\documentclass[fleqn,10pt]{wlscirep}
\usepackage[utf8]{inputenc}
\usepackage[T1]{fontenc}
\usepackage{algorithm}
\usepackage{algpseudocode}
\usepackage[algo2e]{algorithm2e}
\usepackage{bm}

\usepackage{subcaption} 
\usepackage{arydshln}
\usepackage{booktabs}  
\usepackage{graphicx}
\usepackage{color, xcolor}
\usepackage{colortbl}
\usepackage{longtable}
\usepackage{tabularx} 
\usepackage{float} 
\usepackage{array}  
\usepackage[numbers]{natbib} 
\usepackage{multirow}

\usepackage{tikz}

\title{PharmAgents: Building a Virtual Pharma with Large Language Model Agents}

\author[1,2,*]{Bowen Gao}
\author[3,*]{Yanwen Huang}
\author[3,*]{Yiqiao Liu}
\author[4,*]{Wenxuan Xie}
\author[1]{Wei-Ying Ma}
\author[1]{Ya-Qin Zhang}
\author[1,$\dagger$]{Yanyan Lan}

\affil[1]{Institute for AI Industry Research (AIR), Tsinghua University}
\affil[2]{Department of Computer Science and Technology, Tsinghua University}
\affil[3]{Department of Pharmaceutical Science, Peking University}
\affil[4]{School of Future Technology, South China University of Technology}
\affil[*]{Equal Contributions}
\affil[$\dagger$]{Corresponding author: lanyanyan@air.tsinghua.edu.cn}

\begin{abstract}
The discovery of novel small molecule drugs remains a critical scientific challenge with far-reaching implications for treating diseases and advancing human health. Traditional drug development—especially for small molecule therapeutics—is a highly complex, resource-intensive, and time-consuming process that requires multidisciplinary collaboration. Recent breakthroughs in artificial intelligence (AI), particularly the rise of large language models (LLMs), present a transformative opportunity to streamline and accelerate this process. In this paper, we introduce PharmAgents, a virtual pharmaceutical ecosystem driven by LLM-based multi-agent collaboration. PharmAgents simulates the full drug discovery workflow—from target discovery to preclinical evaluation—by integrating explainable, LLM-driven agents equipped with specialized machine learning models and computational tools. Through structured knowledge exchange and automated optimization, PharmAgents identifies potential therapeutic targets, discovers promising lead compounds, enhances binding affinity and key molecular properties, and performs in silico analyses of toxicity and synthetic feasibility. Additionally, the system supports interpretability, agent interaction, and self-evolvement, enabling it to refine future drug designs based on prior experience. By showcasing the potential of LLM-powered multi-agent systems in drug discovery, this work establishes a new paradigm for autonomous, explainable, and scalable pharmaceutical research, with future extensions toward comprehensive drug lifecycle management.
\end{abstract}
\begin{document}

\flushbottom
\maketitle

\thispagestyle{empty}

\section{Introduction}

\begin{figure}[h]
\begin{center}
\centerline{\includegraphics[width=1\columnwidth]{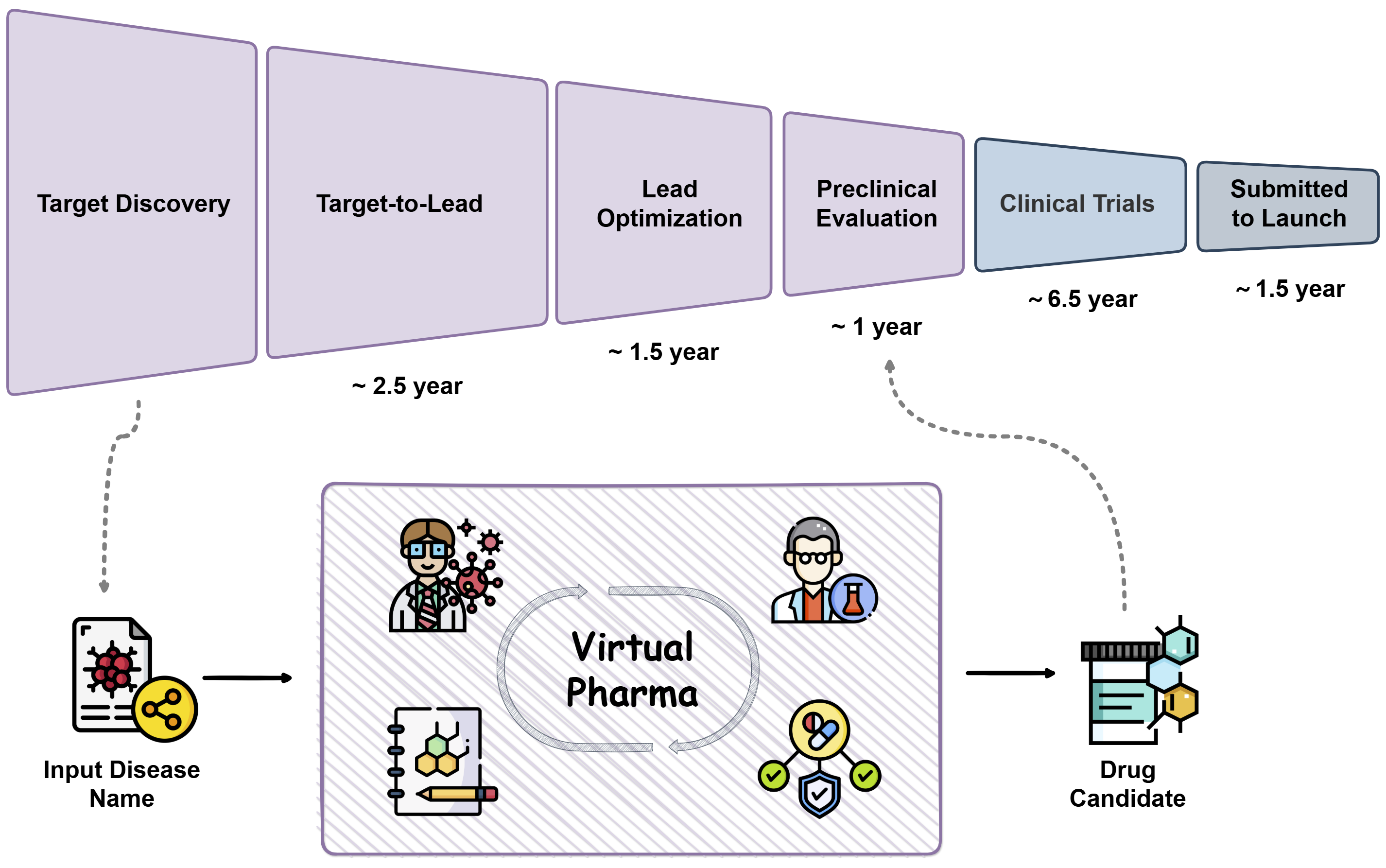}}
\caption{The Virual Pharma (PharmAgents) simulates the drug discovery process from target discovery to preclinical evaluation.}
\label{fig: virtual}
\end{center}
\end{figure}

In the pharmaceutical industry, drug discovery is driven by the imperative to address unmet medical needs by producing safe, effective, and quality-controlled therapeutics that ultimately benefit patients while also yielding social and financial returns. Typically, the targeted small-molecule drug discovery process involves identifying and validating a disease-relevant biological target, screening large chemical libraries to find promising initial chemical compounds, refining these compound structures into optimized leads, and then advancing candidates through extensive preclinical and clinical evaluations. Given the complexity of this process, drug discovery demands the expertise of interdisciplinary teams, combining knowledge from biology, chemistry, pharmacology, and data science, which limits the pharmaceutical exploration of a wide range of diseases. Although advancements such as high-throughput screening~\cite{bajorath2002integration,mayr2009novel} and computational modeling~\cite{hassan2016computer,macalino2015role} have introduced greater efficiencies, the industry continues to face high attrition rates and escalating development costs. These challenges underscore the urgent need for a more intelligent and scalable solution—one that integrates cross-disciplinary knowledge, automates key stages of the drug discovery process, and enables seamless collaboration between specialized expertise to reduce time, cost, and risk, ultimately driving a more efficient and effective path to therapeutic development.


Recent advancements in artificial intelligence (AI) and machine learning (ML) have unlocked new possibilities for accelerating drug discovery. A diverse array of ML models has emerged, facilitating tasks such as virtual screening \cite{gao2023drugclip,zhou2024s}, structure-based molecule generation \cite{peng2022pocket2mol,guan3d, guan2023decompdiff, qumolcraft}, binding affinity prediction\cite{ozturk2018deepdta, zhang2023planet, jones2021improved}, toxicity assessment\cite{cavasotto_machine_2022, sharma_accurate_2023}, and synthesis pathway design\cite{chen_generalized-template-based_2022, zeng2024ualignpushinglimittemplatefree}. While these models have demonstrated impressive accuracy and efficiency in expediting various stages of drug discovery, they often operate independently, lacking the seamless integration required for real-world pharmaceutical workflows. Moreover, a critical limitation lies in their lack of explainability, which hinders full automation and diminishes trust among pharmaceutical experts. Explainability is especially crucial in the early stages of autonomous AI-driven drug discovery, where human oversight is essential to validate model decisions and ensure safety. Without transparency and interpretability, these AI models may struggle to gain acceptance in clinical and regulatory environments, ultimately slowing down the transition toward fully autonomous systems. Bridging this gap between performance and explainability is key to realizing the true potential of AI in revolutionizing drug discovery.
Large language model (LLM) agents, equipped with diversified domain knowledge and capable of seamless communication in human language, are an ideal choice for enabling autonomous drug discovery. Multi-agent systems powered by LLMs—where multiple AI agents collaborate using specialized tools—have already demonstrated success across diverse fields, including software development \cite{hongmetagpt}, medical diagnosis \cite{li2024agent}, and autonomous driving \cite{wei2024editable}. These systems showcase how complex tasks can be effectively decomposed and addressed through coordinated agent collaboration, with each agent specializing in a specific role.

In this article, we introduce PharmAgents, a novel paradigm that harnesses the power of LLMs and multi-agent systems to transform the small-molecule drug discovery process. Drawing inspiration from MetaGPT \cite{hongmetagpt}, which structures software development through well-defined roles such as project manager, developer, and quality assurance engineer, PharmAgents adopts a similar role-based strategy to streamline and automate the pharmaceutical research pipeline. Specifically, the complex and iterative drug discovery workflow is decomposed into four key stages: target discovery, lead identification, lead optimization, and preclinical evaluation. For each stage, we design specialized agents—powered by state-of-the-art LLMs and augmented with advanced machine learning models and domain-specific computational tools—that are assigned clearly defined responsibilities aligned with real-world pharmaceutical tasks. These agents are embedded within a structured, collaborative framework that maximizes individual capabilities while enabling synergistic interactions across the system. The LLMs provide broad knowledge, reasoning abilities, and generative capabilities, while the supporting ML models and computational platforms offer high-precision predictions, thus compensating for the inherent limitations of LLMs in areas such as understanding the 3D structures. Each agent is guided by carefully engineered, domain-informed prompts and integrated into a well-defined workflow with clear task decomposition, ensuring coherent execution and collaboration. This cohesive, explainable system significantly enhances efficiency, reliability, and interpretability across the pipeline, paving the way for a new era of AI-driven, autonomous drug discovery.

Experimental results demonstrate that PharmAgents enables rational, transparent, and high-performance decision-making across the entire drug discovery pipeline. Its target discovery module effectively identifies disease-relevant protein targets, selects optimal structural conformations, and accurately locates ligand-binding sites based on molecular properties without error accumulation, even in the context of complex and noisy diseases. For one random tested disease, 16 out of 18 targets identified by the module were marked as appropriate by human experts following thorough investigation. The molecule generation and optimization modules exhibit the ability to design compounds with seemingly contradictory properties for the same target across different disease contexts, capturing disease-specific nuances. These modules consistently outperform state-of-the-art approaches by significantly enhancing key drug development metrics, boosting overall success rates from 15.72\% to 37.94\%. The preclinical evaluation module adds further robustness by accurately assessing metabolism and toxicity, maintaining a low underestimation risk of just 12\%. It also evaluates molecular synthesizability in a manner that not only aligns well with quantitative metrics (Pearson correlation of 0.645) but also delivers clear and interpretable rationale. PharmAgents also exhibits self-evolving capabilities by summarizing and learning from past experiences to refine future outputs, resulting in an increase in success rates from 30\% to 36\% when prior experience is incorporated, With built-in interpretability, inter-agent collaboration, and evolvment capabilities, PharmAgents represents a transformative step toward realizing a fully autonomous, virtual pharmaceutical enterprise that rivals traditional pharma in both innovation and speed.

\section{Preliminaries}

With the advancement of machine learning, numerous models have been applied to small molecule drug discovery, aiming to accelerate the overall process. Common tasks include binding affinity prediction, which seeks to estimate the interaction strength between a protein pocket and small molecules \cite{ozturk2018deepdta, zhang2023planet, jones2021improved}; virtual screening, where models are used to identify potential lead compounds from large molecular libraries \cite{gao2023drugclip,zhou2024s}; and structure-based molecule generation, which involves learning to generate de novo molecules conditioned on the structure of the target protein pocket \cite{peng2022pocket2mol, guan3d, qumolcraft, guan2023decompdiff}. Moreover, for the special needs in chemical synthesis, there are also reaction-based and synthesis-based models which can provide convincing predictions and instructions to chemists from a methodology dimension, either along the forward direction of synthetic route or in the reverse \cite{chen_generalized-template-based_2022, zeng2024ualignpushinglimittemplatefree}. And lastly, some toxicity prediction models are also available nowadays, which are used to tackle various toxicity-related problems, for instance, binary toxicity judgement, toxicity classification and acute toxicity $LD_{50}$  \cite{cavasotto_machine_2022, sharma_accurate_2023}.

LLMs have also been increasingly applied in the fields of drug discovery and chemistry. Some works focus on leveraging LLMs for prediction tasks in small molecule drug discovery \cite{qian2023can, liu2024moleculargpt}, while others explore their use in molecular generation and design \cite{liu2024conversational}. Systems like ChemCrow \cite{bran2023chemcrow} and ChemAgent \cite{tang2025chemagent} integrate LLMs with external tools to perform a variety of chemistry-related tasks. However, the application of LLMs within a multi-agent framework to simulate the entire drug discovery process remains largely unexplored.

\section{Virtual Pharma Framework Design}

\definecolor{darkyellow}{rgb}{0.8, 0.6, 0.02}
\definecolor{darkpink}{rgb}{1, 0.28, 0.58}
\definecolor{darkgreen}{rgb}{0, 0.55, 0}

\begin{figure}[t]
\begin{center}
\centerline{\includegraphics[width=0.9\columnwidth]{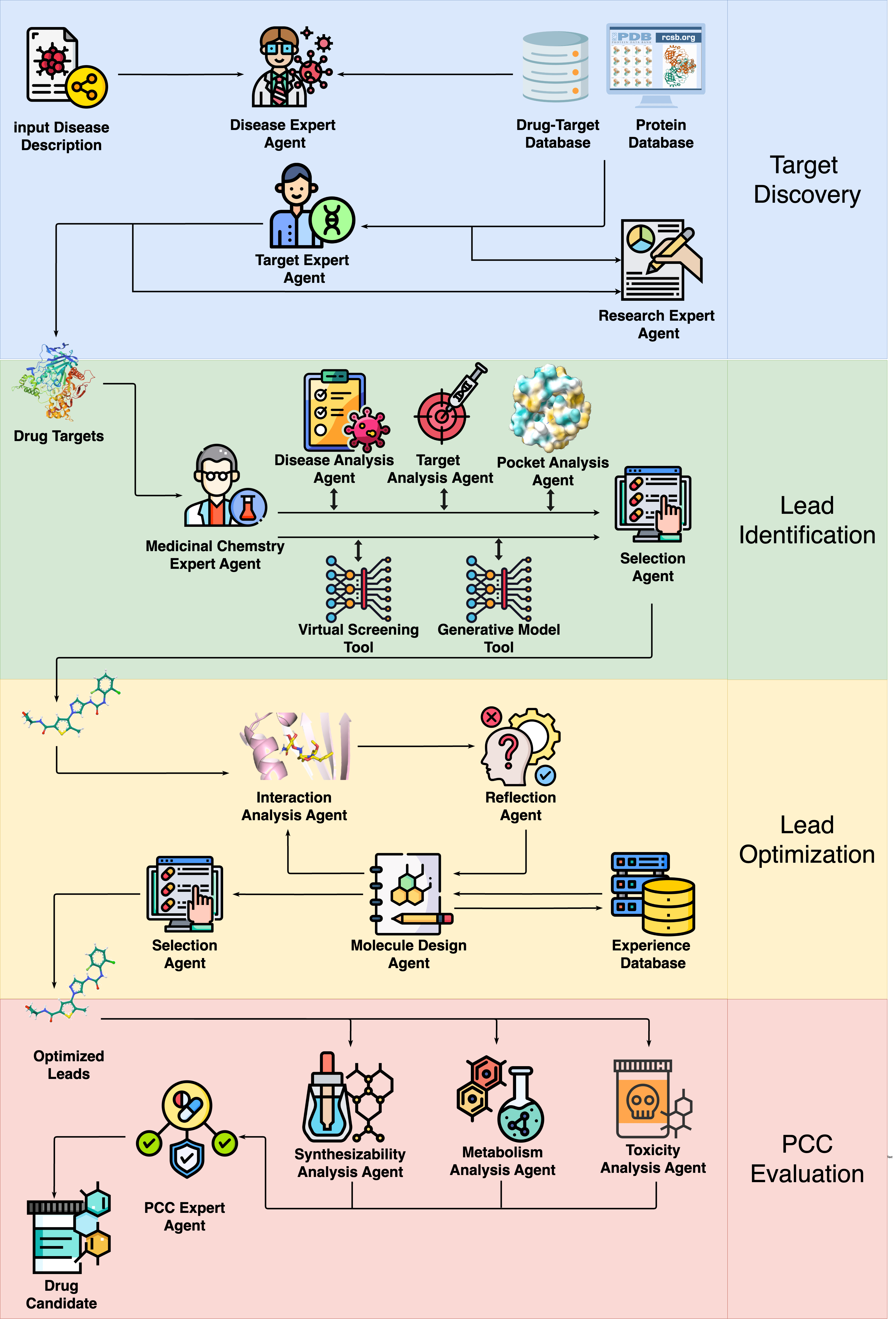}}
\caption{Overall workflow of the PharmAgents.}
\label{fig: whole process}
\end{center}
\end{figure}

The \textbf{Virtual Pharma} framework closely mimics real-world pharmaceutical processes, encompassing the entire drug discovery pipeline—from identifying potential targets for a given disease to lead molecule discovery, optimization, and preclinical evaluation. As illustrated in Figure~\ref{fig: whole process}, we present the complete workflow of \textbf{PharmAgents}. To replicate this pipeline, our Virtual Pharma system is structured into four interconnected modules: \\

\textcolor{blue}{Target Discovery}: Given a disease description provided by the user, this module identifies and outputs potential therapeutic targets associated with the disease. \\

\textcolor{darkgreen}{Lead Identification}: Based on the targets identified in the previous module, this stage analyzes disease context, target information, and pocket structures to generate potential lead compounds. \\

\textcolor{darkyellow}{Lead Optimization}: Using the lead compounds generated in the previous stage, this module refines and optimizes the molecules to enhance their binding affinity and drug-likeness properties. \\

\textcolor{darkpink}{Preclinical candidate (PCC) Evaluation}: This module evaluates the designed molecules for metabolism, toxicity, and synthetic feasibility, providing recommendations on their suitability for further experimental validation. \\

Each module consists of multiple tasks, with each task supported by an \textbf{LLM-based agent} equipped with specialized domain-specific tools or models. These agents work collaboratively to facilitate the entire drug development process. By integrating Large Language Models into every stage, our framework ensures that insights are not only data-driven but also scientifically interpretable, allowing researchers to make well-informed decisions at each step. This approach enhances the \textbf{transparency, interpretability, and efficiency} of drug discovery while maintaining alignment with industry standards.

In the following sections, we detail each module of the framework.







\clearpage

\subsection{Target Discovery}

\begin{figure}[h]
\begin{center}
\centerline{\includegraphics[width=1\columnwidth]{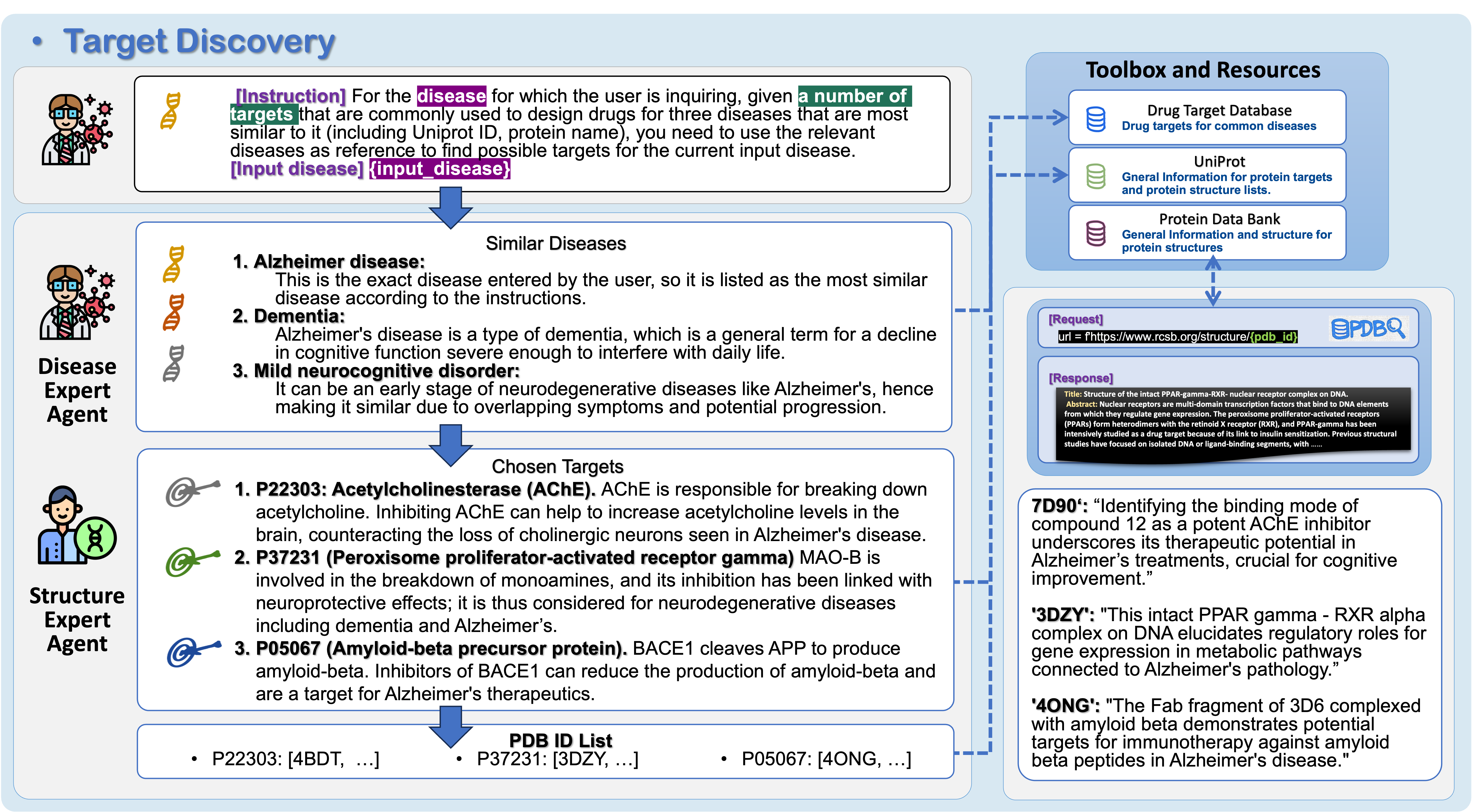}}
\caption{Workflow and example outputs of Target Discovery module.}
\label{fig: target discovery}
\end{center}
\end{figure}





The initial step in drug discovery is the identification of potential targets, which lays the foundation for the entire process. In our approach, we leverage both the knowledge encoded in LLM parameters and information from external databases to support target identification. Upon receiving a disease name from the user, a disease expert determines the relevant condition and retrieves associated drug target data, including potential UniProt IDs~\cite{pundir2017uniprot}. A structure expert then analyzes and filters the corresponding PDB IDs~\cite{berman2000protein} using relevant literature and ligand information. The module ultimately outputs a selected list of PDB IDs along with the chosen pocket center for each PDB, representing the potential target structures for subsequent drug design. This workflow is illustrated in Figure~\ref{fig: target discovery}.

The training corpus of LLMs comprises an extensive collection of foundational research literature and a broad spectrum of domain-specific knowledge. Embedded within these corpora are numerous knowledge associations that often remain unexplored by human experts. Consequently, LLMs possess the potential to uncover implicit patterns and relationships, rendering them well-suited for the task of target identification. When integrated with appropriate tools and databases, and guided by precise task formulations, LLMs can be effectively adapted into domain-specific experts.

The first expert is the \textbf{Disease Expert}. This expert has access to both the \textbf{Drug Target Database} and the \textbf{Unified Protein Database} (\textbf{UniProt}~\cite{pundir2017uniprot}). The former is a database we modified based on the Therapeutic Target Database~\cite{zhou2024ttd}, containing 427 diseases and 1789 corresponding targets. To begin, given a disease under investigation, the expert identifies the three most pathologically similar common diseases in the Drug Target Database. Next, the expert retrieves drug target data for these three reference diseases, including their corresponding UniProt IDs and protein names. By analyzing the mechanisms that link these diseases and their targets, and drawing upon stored knowledge, the expert identifies potential targets (UniProt IDs) for the disease under investigation. Using these UniProt IDs, the expert can then access the general information for each protein target and the corresponding protein structure list from UniProt.

After obtaining the PDB ID list, the \textbf{Structure Expert} will retrieve information about each crystal structure from the \textbf{RCSB Protein Data Bank} (\textbf{RCSB PDB}~\cite{berman2000protein}), including a brief description, abstracts of relevant papers, and a co-crystal ligand name. The co-crystal ligand name, in particular, serves as a key criterion for selection and is crucial in identifying potential targets. Sometimes, a single PDB ID in the database may correspond to multiple ligand names. In these cases, experts must carefully choose the most suitable ligand name based on the structural features. The following step is the final step of this module: PDB filtering. Due to the large number of PDB candidates to be filtered, we employed three strategies to improve the accuracy and robustness of the filtering results, namely \textbf{Grouping Filtering}, \textbf{Reverse Order Consistency}, and \textbf{UniProt Filtering}.

\paragraph{Group Filtering}
In this strategy, the PDB candidates are divided into groups of 100. Within each group, the top 10 candidates are selected based on their potential to bind to the target. After processing all groups, the top 10 candidates from all groups are combined and a final selection of the top 10 is made.

\paragraph{Reverce Order Consistency}
To minimize the impact of the sequence order on the final results, this strategy applies reverse order consistency. After combining all the results from the group filtering step, the entire set of PDB candidates is reversed. The top 10 candidates are selected again from both the original and reversed sets. The final selection is the intersection of the top 10 from both orders. If no intersection is found, the first candidate from each order is selected.

\paragraph{UniProt Filtering}
An additional step is applied to ensure that for each UniProt ID, no more than three PDB IDs are selected. This step is crucial for maintaining the diversity of the predicted targets. This strategy effectively reduces the risk of overrepresentation of any single target, thereby promoting a more balanced and heterogeneous set of potential targets. 

\paragraph{Research Report}
After selecting the PDB targets, all the reasoning trajectories from this process are submitted to a \textbf{Research expert}, who investigates the latest developments in the understanding of the disease and then produces a report based on the target screening process.

\begin{algorithm}[h]
\caption{PDB Filtering for Target Discovery}
\KwIn{PDB candidates (\textit{PDB\_candidates}), Disease (\textit{D})}
\KwOut{Final selected PDBs (\textit{PDB\_final})}

\textbf{Each PDB candidate contains:} \textit{description}, \textit{abstract}, and \textit{co-crystal ligand name}. \\
These are used by Structure Expert in the selection process to identify potential targets for Disease $D$.

\ForEach{group $G_i$ of 100 in \textit{PDB\_candidates}}{
    Reverse the group: $G_i' \gets \text{reverse}(G_i)$\\
    $Top10_{\text{forward}} \gets \text{StructureExpert.select\_top\_10}(G_i, D)$\\
    $Top10_{\text{reversed}} \gets \text{StructureExpert.select\_top\_10}(G_i', D)$\\
    $T_i \gets \text{intersection}(Top10_{\text{forward}}, Top10_{\text{reversed}})$\\
    \If{$T_i$ is empty}{
        $T_i \gets [Top10_{\text{forward}}[0], Top10_{\text{reversed}}[0]]$
    }
    $AllTop10_{\text{groups}} \gets AllTop10_{\text{groups}} \cup T_i$
}

\ \ $T \gets \bigcup AllTop10_{\text{groups}}$\\
$T' \gets \text{reverse}(T)$\\
$Top10_{\text{final\_forward}} \gets \text{StructureExpert.select\_top\_10}(T, D)$\\
$Top10_{\text{final\_reversed}} \gets \text{StructureExpert.select\_top\_10}(T', D)$\\
$T_{\text{final}} \gets \text{intersection}(Top10_{\text{final\_forward}}, Top10_{\text{final\_reversed}})$\\
\If{$T_{\text{final}}$ is empty}{
    $T_{\text{final}} \gets [Top10_{\text{final\_forward}}[0], Top10_{\text{final\_reversed}}[0]]$
}

\textbf{Uniport Filtering:}\\
$PDB_{\text{final}} \gets \text{Uniport\_filter}(T_{\text{final}})$\\
Where $Uniport\_filter$ ensures at most 3 PDBs are selected per \textit{uniport\_id} to enhance the diversity of predicted targets.\\
\KwRet{$PDB_{\text{final}}$}
\end{algorithm}

\subsection{Lead Identification}

After identifying a potential target, the next step is to find potential lead compound for that target. In the laboratory, this can be achieved either by designing new molecules with the expertise of chemists or by performing virtual screening to find promising candidates. In our Virtual Pharma platform, this process begins with LLM agents analyzing the disease, the target information and the pocket structure to generate property requirements, then the agent would use the generative model, the virtual screening model and generate by its on based on the generated requirements. Then the selection agent would use the requirements to select molecules from the three different sources.

The generation agent, primarily powered by LLMs, utilizes several advanced tools that are seamlessly integrated into the platform. These tools assist in generating molecules that are likely to interact with the identified target. One such tool is \textbf{DecompDiff}, a state-of-the-art structure-based drug design model, which is regarded as one of the best for designing molecules based on the target structure. Another crucial tool is \textbf{DrugCLIP}, a leading virtual screening model that efficiently scans vast molecular libraries to identify molecules that are most likely to bind with the target.

In addition to these tools, we also employ several analysis agents that are powered by LLMs. These agents provide critical insights to further refine the generation process. The \textbf{Disease Analysis} agent, for example, analyzes the properties that a molecule should possess in order to become an effective drug for a given disease, based on the disease name. The \textbf{Target Analysis Agent} uses PDB information to retrieve relevant research papers, processing the abstracts of these papers through a language model to identify the characteristics of potential lead compounds for a given target. The \textbf{Pocket Analysis Agent} uses the atom types and 3D coordinates of the target protein pocket stored in the .pdb file to analyze the molecular fragments that are most likely to interact with the pocket.

Additionally, we leverage another generation method that involves the LLM itself, which generates novel molecules based on the insights derived from the previous analyses. This method simulates how expert chemists design drugs based on their accumulated knowledge and understanding of the disease, target, and molecular interactions.

In summary, we have three distinct pipelines for generating potential lead compounds: structure-based design, virtual screening, and LLM-guided molecule generation. After the generation phase, the molecules undergo a selection process, also powered by LLMs, which uses the requirements generated from previous analyses to select the final set of \textit{N} molecules.

\begin{figure}[h]
\begin{center}
\centerline{\includegraphics[width=\columnwidth]{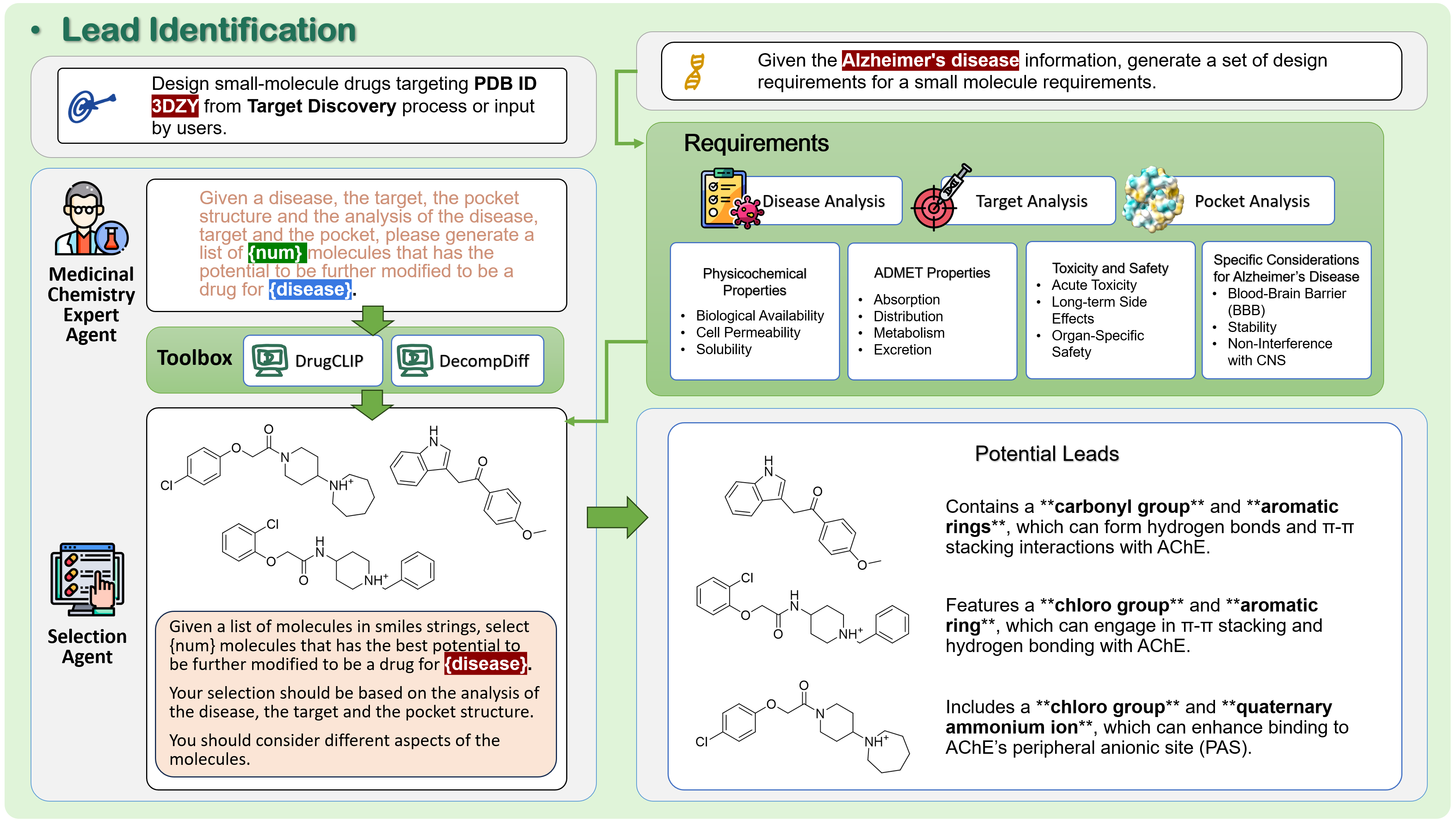}}
\caption{Workflow and example outputs from Lead Identification module.}
\label{fig: lead identification}
\end{center}
\end{figure}

An illustration of the workflow for this module is shown in \textbf{Figure \ref{fig: lead identification}}.

\subsection{Lead Optimization}

\begin{figure}[h]
\begin{center}
\centerline{\includegraphics[width=1\columnwidth]{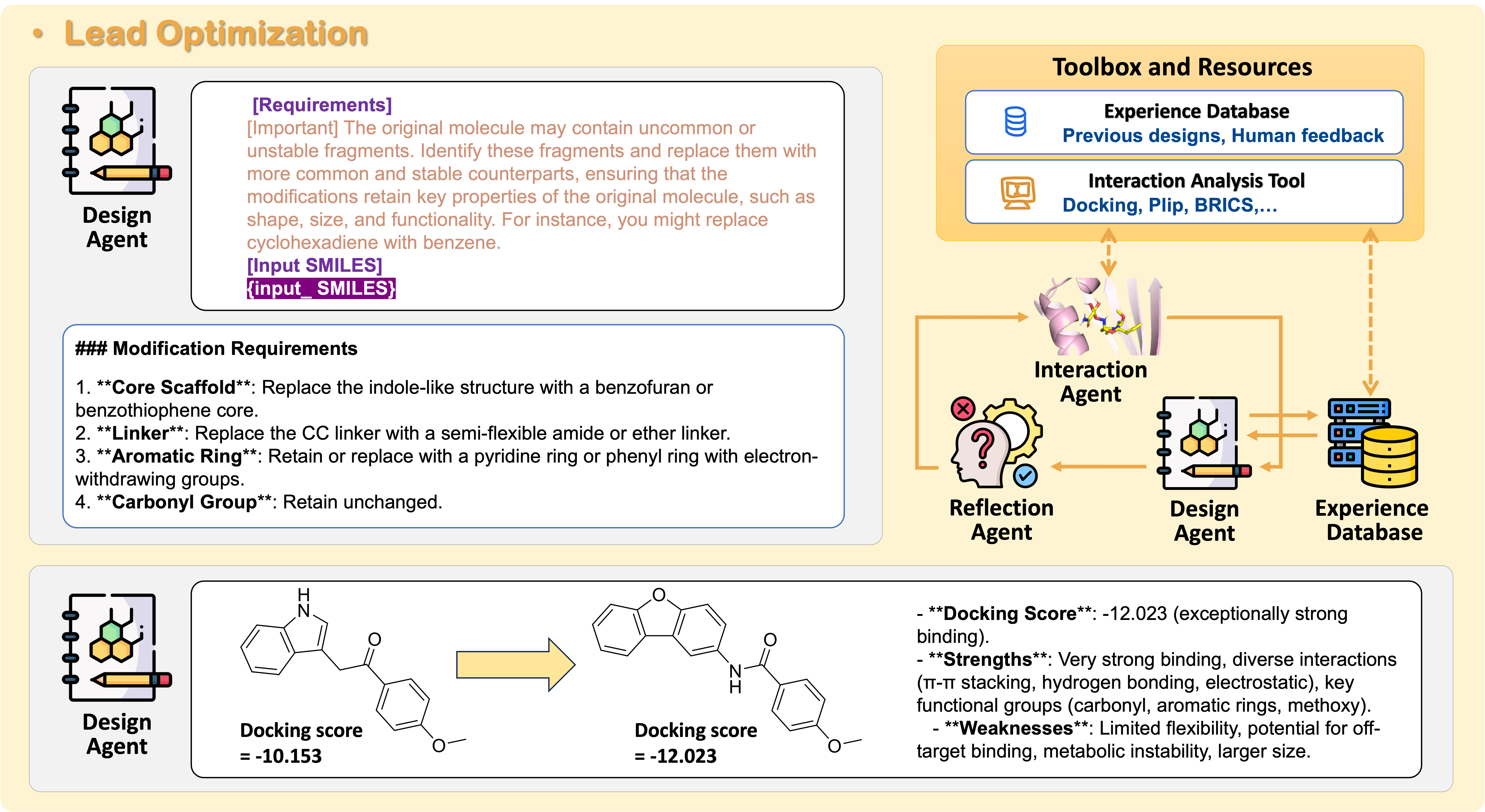}}
\caption{Workflow and example outputs of Lead Optimization module.}
\label{fig: lead opt}
\end{center}
\end{figure}

After generating potential lead compounds, the next critical step in the drug discovery pipeline is lead optimization—a process essential for transforming promising molecules into viable drug candidates. In this section, we introduce the Lead Optimization Module, a key component of our Virtual Pharma framework. Traditionally, lead optimization involves systematically modifying candidate molecules to improve their binding affinity, drug-likeness, synthetic feasibility, and overall structural integrity, while minimizing off-target effects and toxicity. To replicate and enhance this process, our system employs an LLM-based multi-agent architecture that mirrors the iterative nature of expert-driven optimization. This module operates in a closed-loop cycle involving a design agent, an interaction analysis agent, and a reflection agent, which collaboratively propose candidate molecules. The resulting optimized compounds are then assessed by a selection agent, which prioritizes molecules based on a comprehensive evaluation of their pharmacological and chemical properties.

The molecule optimization module consists of several LLM agents working in tandem. The molecules selected in the previous step first undergo analysis by the \textbf{Interaction Analysis Agent}. This agent utilizes two tools: a docking software that docks the molecule to the target to generate the complex structure, and PLIP, a tool that provides detailed information on all non-covalent interactions within the complex. The interaction data from these tools are then used by the LLM agent to analyze the binding interactions between the molecule and the protein pocket.

Next, the interaction data is passed to the \textbf{Design Agent}, which generates suggestions for modifying the original molecule. The goal is to improve the binding affinity, structural reasonability, and druglikeness, ensuring the molecule has favorable drug-like properties. Importantly, the \textbf{Design Agent} does not generate new molecules by itself. Instead, it works in conjunction with the \textbf{Generation Agent}, which carries out the modifications and generates the new, optimized molecule.

Once the new molecule is generated, the \textbf{Interaction Analysis Agents} are re-engaged to assess the binding interactions of the modified molecule. Following this, a \textbf{Reflection Agent} evaluates the changes made during the optimization process. The Reflection Agent considers the design objectives, the original molecule, the modified molecule, and their respective interaction reports to provide feedback on whether the expected design effect has been met. 

This iterative process takes place over several rounds. Each round begins with the original molecule as input and outputs a modified version of the molecule. The feedback provided by the Reflection Agent from previous rounds is integrated into subsequent rounds, guiding further refinements. This design cycle typically occurs over five rounds, with each round progressively improving the molecule based on accumulated insights.

Finally, a \textbf{Selection Agent} analyzes all the generated molecules and their corresponding interaction reports to select the best candidate. The selection is based on both the quality of the molecular interactions and its drug-related properties.

Notably, PharmAgents incorporates an experience database that records all previous designs and reports, mirroring real-world pharmaceutical practices where past trials are systematically documented. When a new drug design cycle begins, the design agent analyzes and summarizes prior experiences to generate valuable insights. This framework enables LLM agents to self-evolve, continuously improving as more designs and results are accumulated. A detailed illustration of this process is provided in Section \ref{sec: evolvement}.

In summary, the molecule optimization module uses a combination of interaction analysis, design suggestions, and reflections to optimize molecules in a manner that mimics real-world drug development. This process ensures that the final molecule is not only highly effective in binding to the target but also possesses favorable drug-like characteristics and synthetic feasibility.

An illustration of the workflow for these agents is shown in \textbf{Figure \ref{fig: lead opt}}.

\subsection{PCC Evaluation}

\begin{figure}[h]
\begin{center}
\centerline{\includegraphics[width=1\columnwidth]{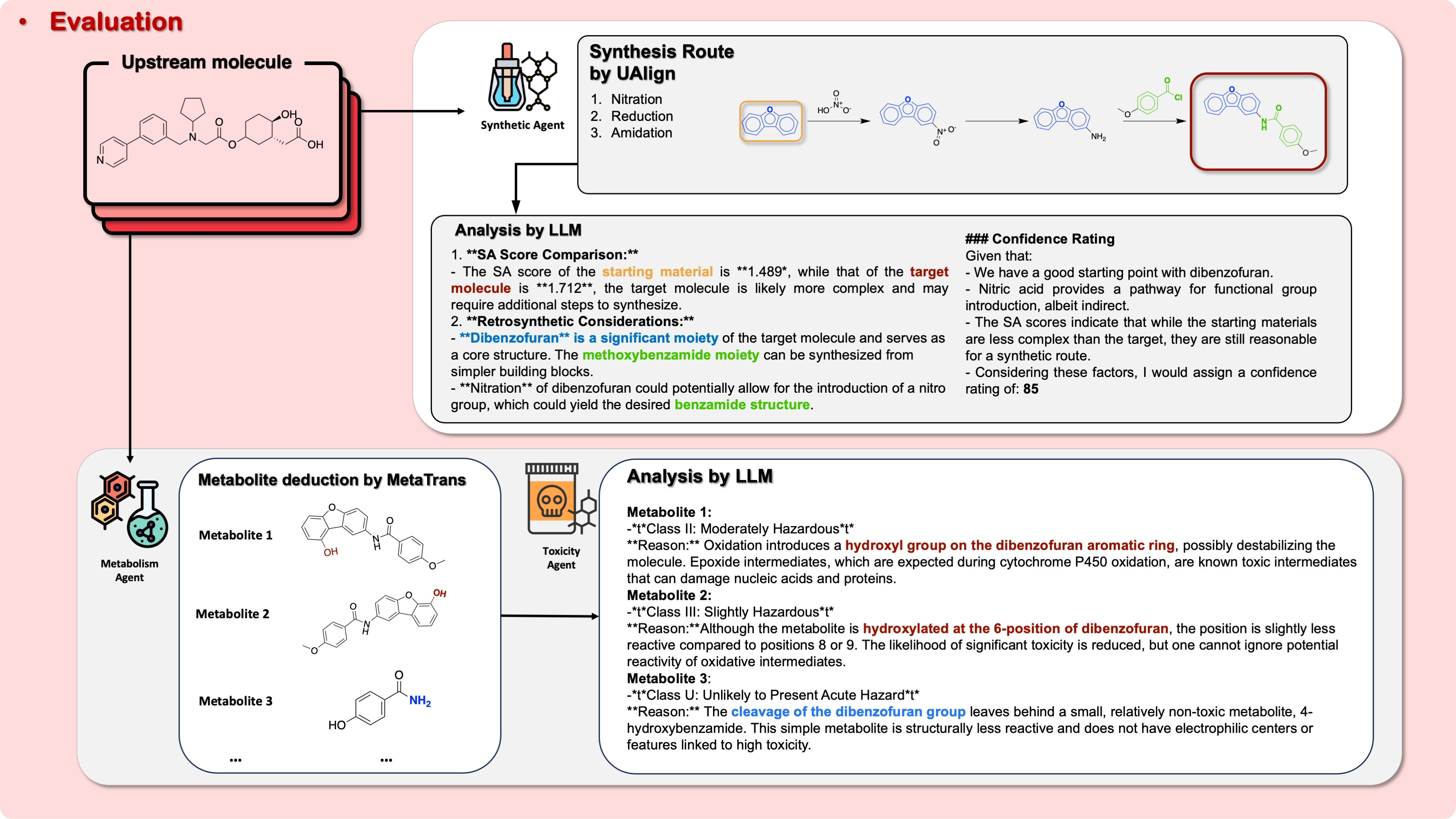}}
\caption{Workflow and example outputs of Preclinical Evaluation module.}
\label{fig: pcc eval}
\end{center}
\end{figure}

In real-life drug development, ensuring the overall curative efficacy of a chemical outweighs its potential risks is fairly important. Also, it is non-trivial to confirm that a molecule's synthetic feasibility can meet the needs in mass-production, particularly for the ones designed through SBDD generative methods. Therefore, we simulate this process by introducing the last component in our virtual pharma as the pre-clinical evaluation module, which functions as a filter to screen the molecules generated from the upstream tasks from the perspectives of toxicity and synthesizability. Similar to the aforementioned modules, this module is also based on multiple LLMs, and augmented by two state-of-the-art Deep Learning models, along with professional knowledge about drug metabolism, drug synthesis and toxicity that LLMs can utilize during their analysis.

On receiving the upstream chemicals, the data will first be fed into the \textbf{Metabolism and Toxicity Assessment Agent}. This agent aims to deduct the metabolites that are most likely to be produced in human internal environment for the given molecules, and then decide whether the original molecule and its derivatives are of any degrees of acute toxicity. This process is achieved under the assistance by a high-performance model named \textbf{MetaTrans}~\cite{metatrans} which implements metabolic modifications to the input molecules and forward them into the LLMs for further inferences. Some brief knowledge of acute toxicity criteria aligning with that of the World Health Organization (WHO) is also involved by pre-insertion into the context of the LLMs. The toxicity prediction logic is based on the assumption that, the more similar two molecules are, the closer their toxicity categories will be. Specifically, we collected a curated database from \textbf{TOXRIC}~\cite{10.1093/nar/gkac1074}, including approximately 10,000 molecules and their LD50 information as the reference data (can be switched to larger and more accurate database to achieve better results), and used their \textbf{Morgan Fingerprint}~\cite{doi:10.1021/ci100050t} of radius 3, which corresponded to \textbf{ECFP6}, as the similarity measurement. When predicting the toxicity of an unknown molecule, we first find the n (default to be 5) most similar molecules from the database with the Tanimoto similarity of the fingerprints higher than 0.2, then integrate their IUPAC name, toxicity categories, etc. into a comprehensive context as the addition of the pre-inserted knowledge, and finally let the LLM reason out its prediction.

Another paralleling agent in this module is called the \textbf{Synthesis Assessment Agent}, whose job is to evaluate the synthetic viability of the upstream molecules. Countless preceding cases have shown that relying only on synthesizability indices is not sufficient, therefore, this agent is equipped with retrosynthesis analysis capabilities to judge the synthesis routes in a step-wise paradigm, by resorting to one of the latest and leading deep learning model in this domain, \textbf{UAlign}~\cite{zeng2024ualignpushinglimittemplatefree}. Since the model was originally developed as a one-step retrosynthesis prediction tool, we additionally designed a selection scheme shown below to sort out a practical synthetic pathway for each molecule, and made sure no cyclic pathways are generated. Later on, the whole pathway is fed into the LLM and summarized as the conversational context, then, the starting precursors corresponding to each upstream molecule are given to the LLM and the synthetic availability of the target product is rated using the confidence of successful synthesis (from 0 to 100) according to the LLM's inherent knowledge and the retrosynthesis knowledge that we inserted.

An illustration of the workflow for this module is shown in \textbf{Figure \ref{fig: pcc eval}}.

\begin{algorithm}[H]
\caption{Synthetic Pathway Selection}
\label{alg:synthetic_path}
\KwIn{Product molecule (\textit{ori\_prd}), Retrosyntheis steps(\textit{steps}), \textit{SA\_Threshold}(default=2.0)}
\KwOut{Synthetic Pathway Inferred from Retrosynthesis Analysis (\textit{OutRXNs}) and the Starting Molecules as Synthetic Precursors (\textit{StartingMols})}

$\textit{prd} \gets \textit{ori\_prd}\\$
\ForEach{step in range(steps)}{
    \If{SAscore(\textit{prd}) is not < SA\_Threshold}{
        $\textit{predictions} \gets \textbf{UAlign}(\textit{prd})\\$
        $\textit{main reactant list} \gets \text{Empty List}\\$
        \ForEach{prediction in predictions}{
            \ForEach{reactant in prediction}{
                $\textit{main reactant} \gets argmax(\text{Tanimoto Similarity}(\text{ECFP6}(\textit{ori\_prd}), \text{ECFP6}(\textit{reactant})))\\$
                Add \textit{main reactant} to \textit{main reactant list}
            }
        }
        Add the \textit{prediction} with lowest MolWeight(\textit{main reactant}) to \textit{OutRXNs}\\
        $\textit{prd} \gets argmin(\text{MolWeight}(\textit{main reactant}))$
    }
    \Else{
        $\textit{StartingMols} \gets \textit{prd}\\$
        \textbf{Exit the loop}
    }
}
\end{algorithm}

For both modules above, every conclusive analysis about the given molecules is included into the same comprehensive report of the molecules respectively, before eventually, all the reports of the whole batch of upstream molecules will be sequentially transmitted to the \textbf{Report Assessment Agent}. This final agent is equipped with all the knowledge mentioned above, and it is intended to work as a binary classification tool to ultimately judge if a molecule has the potential to be further delivered into the clinical trials. For instance, a molecule with low toxicity and high synthesis complexity will be rejected for its low availability despite the desired properties, and vice versa, so that the qualified molecules will be neither too difficult to produce, nor too hazardous as drug candidates.

An illustration of the workflow for these agents is shown in \textbf{Figure~\ref{fig: pcc eval}}.

\section{Results}

\subsection{Results for each module}

\subsubsection{Target Discovery}
Target discovery is the foundation of rational drug design. This process entails identifying a disease-relevant protein and determining its 3D structure. Experts built on LLMs can draw on authoritative databases, alongside their internal biochemical knowledge, to explore potential links between diseases and protein targets, and further with protein complex structures.
Because drug target discovery requires deep biological understanding at multiple levels, LLMs can support it by leveraging the vast amounts of literature, sequence data, and experimental findings they have encountered during training.



To evaluate the capability and robustness of our automated target discovery module, we designed a small test set. Four different models were included: \textbf{GPT-4o}, \textbf{GPT-4o-mini}, \textbf{DeepSeek-V3}, and \textbf{DeepSeek-R1}. Each model received disease inputs at three different levels. As discussed in previous sections, we maintain an extensive disease-target database that covers classical diseases and their corresponding drug targets (identified by UniProt IDs). The LLMs can leverage this database to draw inspiration and uncover potential knowledge associations from existing cases. Based on this, we designed test cases with three samples for each disease group. The first sample is a disease that exists in the database, the second is an alternative name for the first disease (which is not in the database), and the third is a disease that is not included in the database but is similar to the first one. This design allows us to assess the potency of the LLMs on multiple levels, while also testing the robustness of this module. Three sets of examples from the test set are presented in Table \ref{tab:diseases}.

\begin{table}[htbp]
  \centering
  \begin{tabular}{c|p{3.2cm}|p{4.8cm}|p{6.5cm}}
    \toprule
    \textbf{Group ID} & \textbf{Disease in Database} & \textbf{Same Disease, Different Name} & \textbf{Similar Pathology, Different Disease} \\
    \midrule
    1 & Atopic eczema & Atopic dermatitis & Psoriasis \\
    2 & Crohn disease & Regional enteritis & Ulcerative colitis \\
    3 & Sepsis & Septicaemia & Systemic inflammatory response syndrome \\
    \bottomrule
  \end{tabular}
  \caption{Examples from test set of \textbf{Target Discovery}}
  \label{tab:diseases}
\end{table}

We observed that the outputs generated by the different models for the nine diseases varied in formulation but exhibited overall similarity. The correctness of the predicted targets—particularly for diseases in the early stages of research—is inherently subjective and difficult to evaluate quantitatively. Nonetheless, the outputs from all models were considered suitable for subsequent small molecule discovery and design efforts. Expert evaluations by human reviewers confirmed the general plausibility of the predictions. Notably, the Drug Target Database played a decisive role in guiding the target identification process.

Using atopic dermatitis as a representative example, Table~\ref{tab:main_result} summarizes the comparative outputs of different LLM models, including automatically generated UniProt IDs, PDB IDs, and selected ligands used to define the binding pocket, along with corresponding expert evaluations. It is important to emphasize that each stage of this mapping—ranging from disease to protein target, target to protein structure, and structure to pocket-defining ligand—is nontrivial, and error accumulation across steps is difficult to avoid.

Atopic dermatitis, also known as atopic eczema, is a chronic, relapsing inflammatory skin disease affecting approximately 10\% of the global population~\cite{frazier2020atopic}. Its pathogenesis involves a complex interplay between immune dysregulation, epidermal barrier dysfunction—often driven by genetic mutations—and environmental triggers. These factors collectively result in impaired skin barrier integrity, persistent pruritus, and characteristic eczematous lesions~\cite{sroka2021molecular}.

In this case study, we systematically analyzed the accuracy of model-generated mappings. First, the choice of protein targets was found to be accurate. Specifically, three Janus kinase (JAK) isoforms—\textcolor{darkpink}{JAK1 (UniProt ID: P23458)}, JAK2 (O60674), and JAK3 (P52333)—as well as their upstream regulator IL-4R$\alpha$ (P24394), were identified. This selection aligns well with current clinical guidelines, which include the use of targeted biologics such as \textcolor{darkyellow}{dupilumab}, a monoclonal antibody against IL-4R$\alpha$~\cite{blauvelt2017long}, and small-molecule JAK inhibitors including baricitinib (selective for JAK1 and JAK2)\cite{melo2022baricitinib}, \textcolor{darkpink}{abrocitinib (JAK1-selective)~\cite{crowley2020abrocitinib, deeks2021abrocitinib}}, and \textcolor{darkpink}{upadacitinib (JAK1-selective)~\cite{parmentier2018vitro}}, all of which modulate key inflammatory pathways implicated in atopic dermatitis~\cite{tay2016guidelines, chovatiya2021jak}.

Second, we observed that JAK1 was the most frequently selected target, chosen in 7 out of 18 model-generated predictions. This trend is noteworthy, as selective JAK1 inhibitors have shown promise in clinical development. While non-selective JAK inhibitors have demonstrated efficacy and safety, \textcolor{darkpink}{selective JAK1 inhibition may enhance safety by reducing off-target effects, while maintaining therapeutic benefit~\cite{schwartz2017jak, ferreira2020selective}. Their oral availability and favorable tolerability profiles further support their inclusion in emerging treatment algorithms for atopic dermatitis.}

Third, we found that 16 out of the 18 selected protein structures were appropriate. The two incorrect cases (8K4Q and 6WGL), both generated by DeepSeek-R1, involved complex structures of IL-4R$\alpha$ bound to biologic drugs, including the clinically approved \textcolor{darkyellow}{dupilumab} (6WGL). While these structures are not incorrect in a biological sense, the selection of a biologic-bound complex is less suitable for defining small-molecule binding pockets. Nevertheless, this indicates that the overall mapping from disease to structure was generally accurate; the primary limitation lay in ligand selection for pocket definition.

Finally, most of the selected ligands were promising small molecules with potential relevance to atopic dermatitis. Notably, several correspond to clinical or investigational agents with demonstrated or anticipated therapeutic effects. These include next-generation JAK inhibitors, reinforcing the utility of LLM-assisted mapping in identifying therapeutically relevant targets and molecules.

\begin{table}[htbp]
\renewcommand{\arraystretch}{1.2} 
\renewcommand{\familydefault}{\rmdefault}
\resizebox{\textwidth}{!}{
\begin{tabular}{lll>{\columncolor{yellow!2}\small\raggedright\arraybackslash}p{11.5cm}>{\columncolor{blue!2}\small\raggedright\arraybackslash}p{8cm}} 
\hline
\textbf{Models} & \textbf{Uniprot IDs} & \textbf{PDB IDs} & \textbf{Ligands Chosen to Define Pocket} & \textbf{Human Expert Comment} \\
\hline
\multirow{5}{*}[-4ex]{\textbf{GPT-4o-mini}} & O60674 & 5TQ4 & 6-(2-ethyl-4-hydroxyphenyl)-1H-indazole-3-carboxamide & The chosen ligand is with potential for treating lung and \textbf{skin inflammatory diseases}~\cite{jones2017design} \\
& O60674 & 3LPB & N-methyl-4-[3-(3,4,5-trimethoxyphenyl)quinoxalin-5-yl]benzenesulfonamide & - \\
& O60674 & 4AQC & 8-(4-methylsulfonylphenyl)-N-(4-morpholin-4-ylphenyl)-[1,2,4]triazolo[1,5-a]pyridin-2-amine & - \\
& P52333 & 6GLB & 1-phenylurea & - \\
& \textcolor{darkpink}{P23458} & 4IVC & (trans-4-{2-[(1R)-1-hydroxyethyl]imidazo[4,5-d]pyrrolo[2,3-b]pyridin-1(6H)-yl}cyclohexyl)acetonitrile & The chosen ligand is a potent and \textcolor{darkpink}{highly selective JAK1} inhibitor with a scaffold similar to \textbf{upadacitinib}~\cite{zak2013identification}\\
\cdashline{1-5} 
\multirow{4}{*}[-5.5ex]{\textbf{GPT-4o}} & \textcolor{darkpink}{P23458} & 6N7B & N-[3-(5-chloro-2-methoxyphenyl)-1-methyl-1H-pyrazol-4-yl]-1H-pyrazolo[4,3-c]pyridine-7-carboxamide & \textcolor{darkpink}{JAK1} in complex with a highly potent JAK1/2 inhibitor~\cite{zak2019discovery} \\
& \textcolor{darkpink}{P23458} & 4EHZ & 2-methyl-1-(piperidin-4-yl)-1,6-dihydroimidazo[4,5-d]pyrrolo[2,3-b]pyridine & \textcolor{darkpink}{JAK1} in complex with a potent and \textcolor{darkpink}{highly selective JAK1} inhibitor~\cite{zak2012discovery}\\
& \textcolor{darkpink}{P23458} & 6SMB & ~{N}-[3-[2-[(3-methoxy-1-methyl-pyrazol-4-yl)amino]-5-methyl-pyrimidin-4-yl]-1~{H}-indol-7-yl]-2-methyl-pyridine-3-carboxamide & \textcolor{darkpink}{JAK1} in complex with a potent and \textcolor{darkpink}{highly selective JAK1} inhibitor~\cite{su2020discovery} \\
& P52333 & 7C3N & 3-[(3S,4R)-3-methyl-7-(7H-pyrrolo[2,3-d]pyrimidin-4-yl)-1,7-diazaspiro[3.4]octan-1-yl]-3-oxidanylidene-propanenitrile & The chosen ligand \textbf{delgocitinib} is approved  for the treatment of atopic dermatitis~\cite{noji2020discovery, dhillon2020delgocitinib} \\
\cdashline{1-5} 
\multirow{3}{*}[-2ex]{\textbf{DeepSeek-V3}} & P52333 & 4QT1 & 1-{(3S)-1-[(2-methylpropyl)sulfonyl]piperidin-3-yl}-3-(5H-pyrrolo[2,3-b]pyrazin-2-yl)urea & - \\
& P52333 & 5LWM & 1-phenylurea & - \\
& P52333 & 6DB3 & [(1S)-1-methyl-6-(7H-pyrrolo[2,3-d]pyrimidin-4-yl)-2,3-dihydro-1H-inden-1-yl]cyanamide & - \\
\cdashline{1-5} 
\multirow{6}{*}[-5.5ex]{\textbf{DeepSeek-R1}} & P24394 & 8K4Q & 2-acetamido-2-deoxy-beta-D-glucopyranose & A complex of IL-4R$\alpha$ and a humanized nanobody~\cite{zhu2024novel} \\
& P24394 & 6WGL & 2-acetamido-2-deoxy-beta-D-glucopyranose & A complex of IL-4R$\alpha$ extracellular domain and \textcolor{darkyellow}{dupilumab} Fab~\cite{lieu2020rapid} \\
& \textcolor{darkpink}{P23458} & 6DBN & [(1S)-2,2-difluorocyclopropyl][(1R,5S)-3-{2-[(1-methyl-1H-pyrazol-4-yl)amino]pyrimidin-4-yl}-3,8-diazabicyclo[3.2.1]octan-8-yl]methanone & \textcolor{darkpink}{JAK1} in complex a small molecule~\cite{fensome2018dual}\\
& \textcolor{darkpink}{P23458} & 6TPE & 2-[4-(3-methyl-6-oxidanylidene-1,7-dihydropyrazolo[3,4-b]pyridin-4-yl)cyclohexyl]ethanenitrile & \textcolor{darkpink}{JAK1} in complex with a JAK1 inhibitors with \textcolor{darkpink}{excellent subtype selectivity}~\cite{hansen2020fragment} \\
& \textcolor{darkpink}{P23458} & 6SMB & ~{N}-[3-[2-[(3-methoxy-1-methyl-pyrazol-4-yl)amino]-5-methyl-pyrimidin-4-yl]-1~{H}-indol-7-yl]-2-methyl-pyridine-3-carboxamide & \textcolor{darkpink}{JAK1} in complex with a potent and \textcolor{darkpink}{highly selective JAK1} inhibitor~\cite{su2020discovery} \\
& P52333 & 7C3N & 3-[(3S,4R)-3-methyl-7-(7H-pyrrolo[2,3-d]pyrimidin-4-yl)-1,7-diazaspiro[3.4]octan-1-yl]-3-oxidanylidene-propanenitrile & JAK3 in complex with a small molecule with the effect of \textbf{treating inflammatory skin disorders}~\cite{noji2020discovery} \\
\hline
\end{tabular}
}
\caption{
The evaluation result of \textbf{Target Discovery} with different LLMs.
}
\label{tab:main_result}
\end{table}


\subsubsection{Lead Identification/Optimization}

In the Lead Identification step, our system is equipped with advanced machine-learning tools for drug design and screening. In this phase, LLMs play a crucial role in analyzing diseases and targets, generating specific molecular requirements tailored to treating a given disease with a particular target.

\begin{figure}[h]
\begin{center}
\centerline{\includegraphics[width=1\columnwidth]{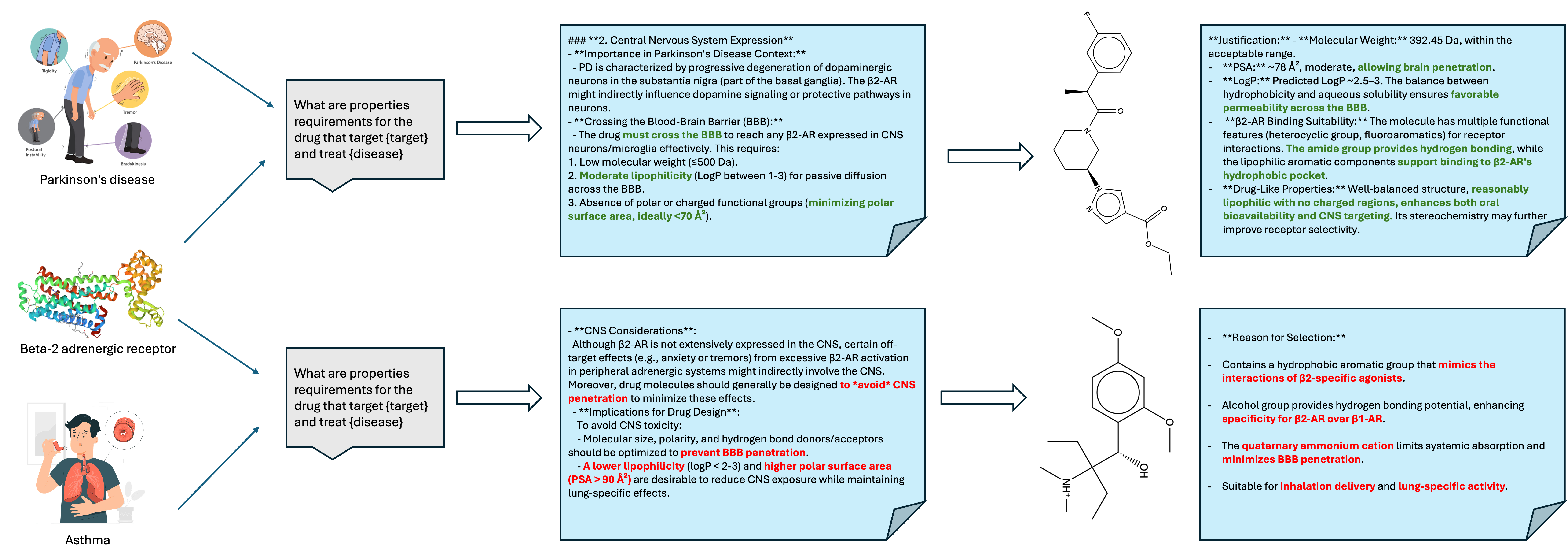}}
\caption{Illustration of prompts and responses for the disease and target analysis agent that leads to different molecules.}
\label{fig: target analysis}
\end{center}
\end{figure}

As illustrated in Figure \ref{fig: target analysis}, when the target remains the same but the disease differs, the analysis agent produces distinct molecular requirements. For instance, in the case of Parkinson’s disease, the molecule must cross the blood-brain barrier (BBB) to be effective in the central nervous system (CNS). Conversely, for asthma, the molecule should avoid BBB penetration to minimize central nervous system effects. Then, based on different generated requirements, various molecules are selected, with the corresponding reasons highlighted in \textcolor{darkgreen}{green} for one disease and \textcolor{red}{red} for another.

\begin{figure}[h]
    \centering

    \begin{subfigure}[b]{0.3\textwidth}
        \centering
        \includegraphics[width=\textwidth]{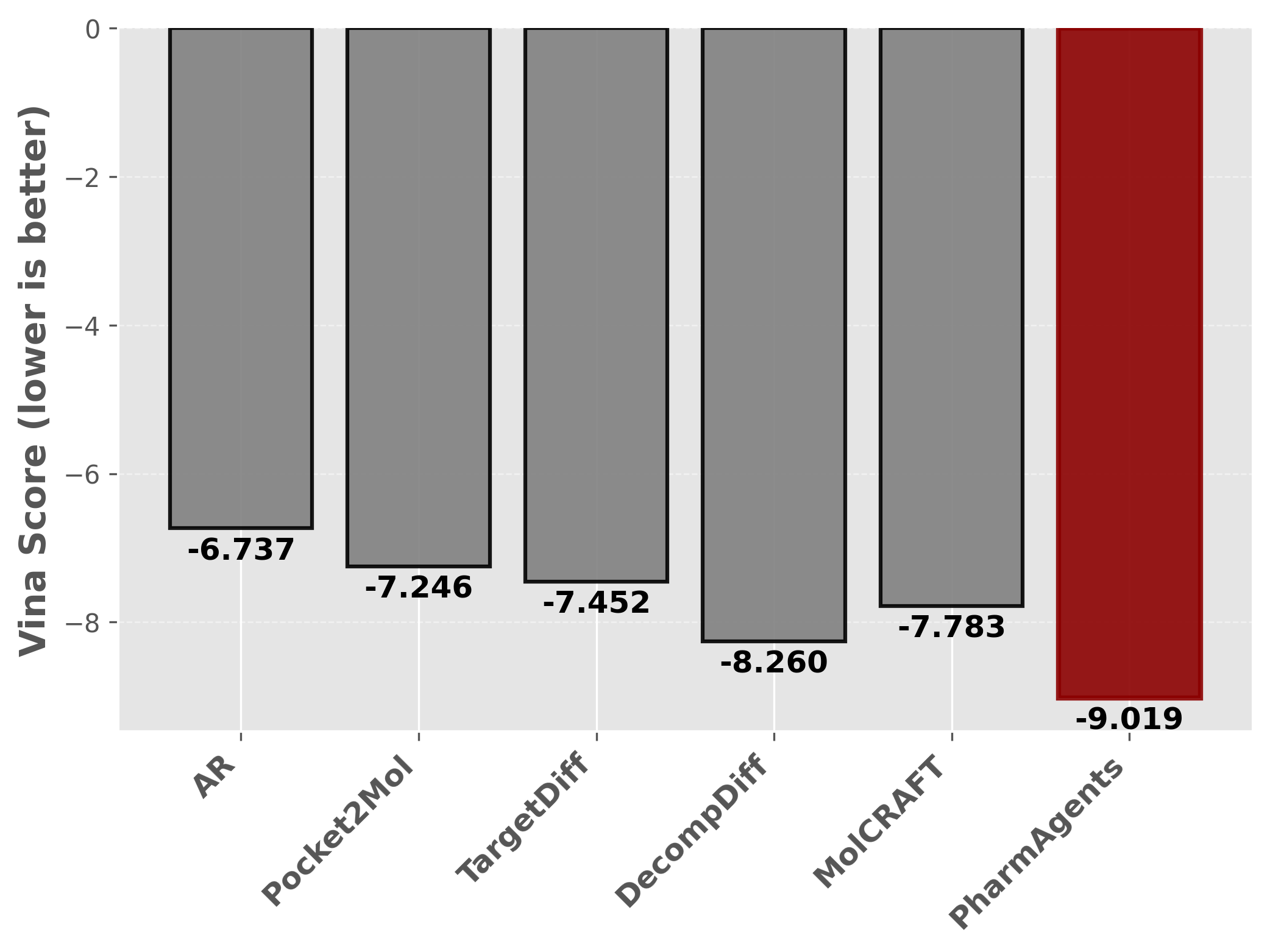}
        \caption{Vina Score (lower is better)}
    \end{subfigure}
    \hfill
    \begin{subfigure}[b]{0.3\textwidth}
        \centering
        \includegraphics[width=\textwidth]{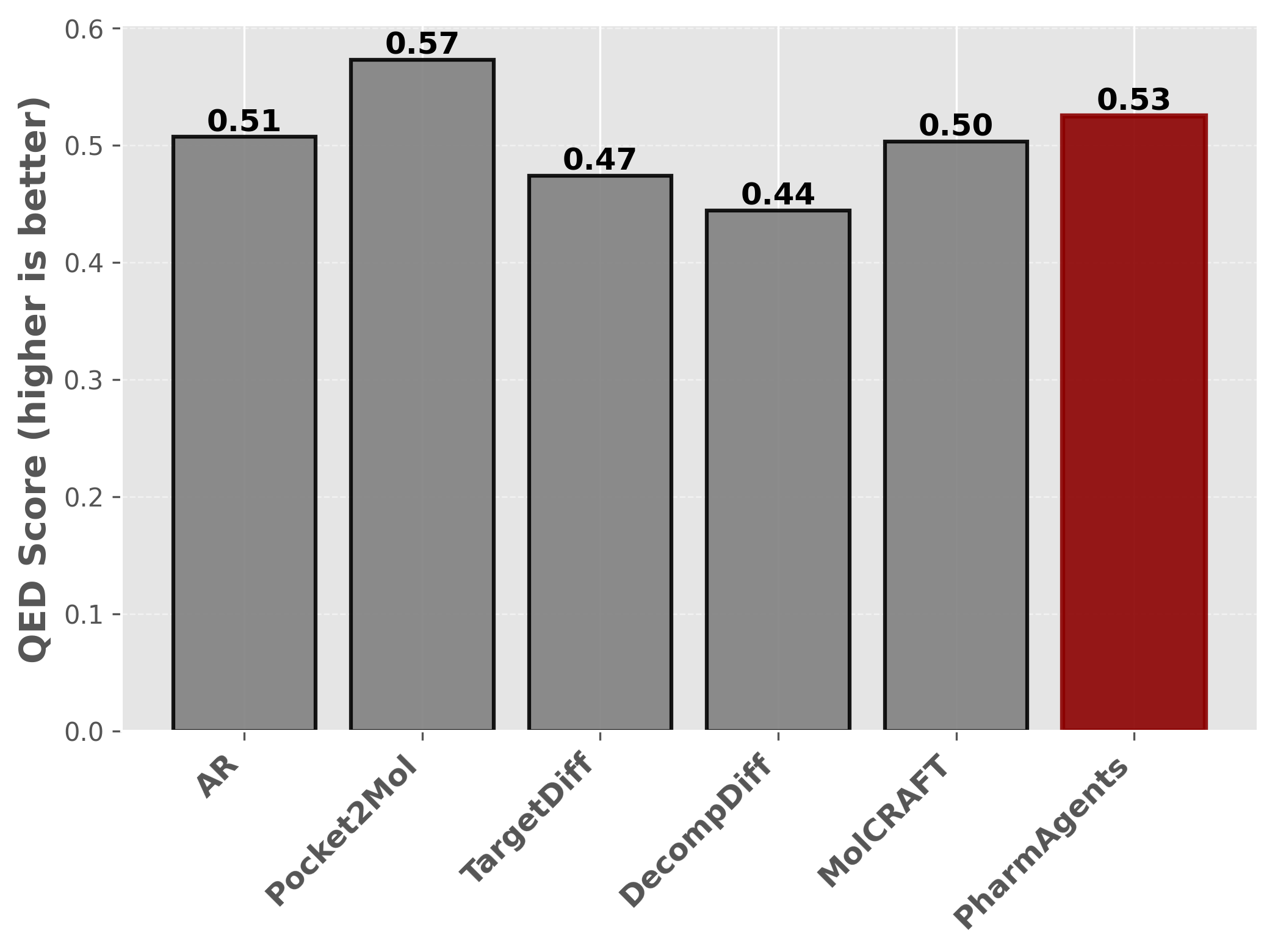}
        \caption{QED Score (higher is better)}
    \end{subfigure}
    \hfill
    \begin{subfigure}[b]{0.3\textwidth}
        \centering
        \includegraphics[width=\textwidth]{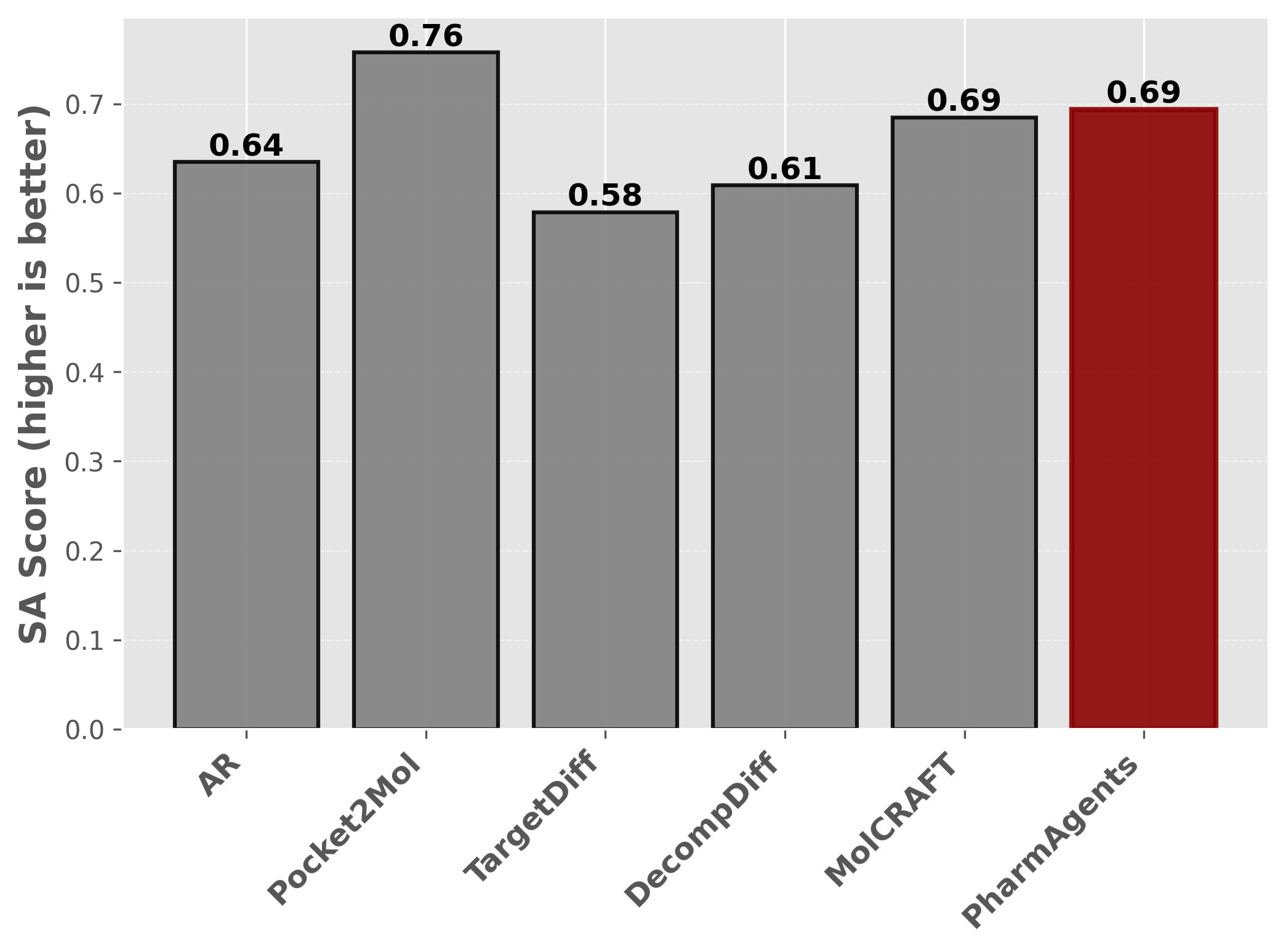}
        \caption{SA Score (higher is better)}
    \end{subfigure}
    
    \vspace{1cm}

    \begin{subfigure}[b]{0.3\textwidth}
        \centering
        \includegraphics[width=\textwidth]{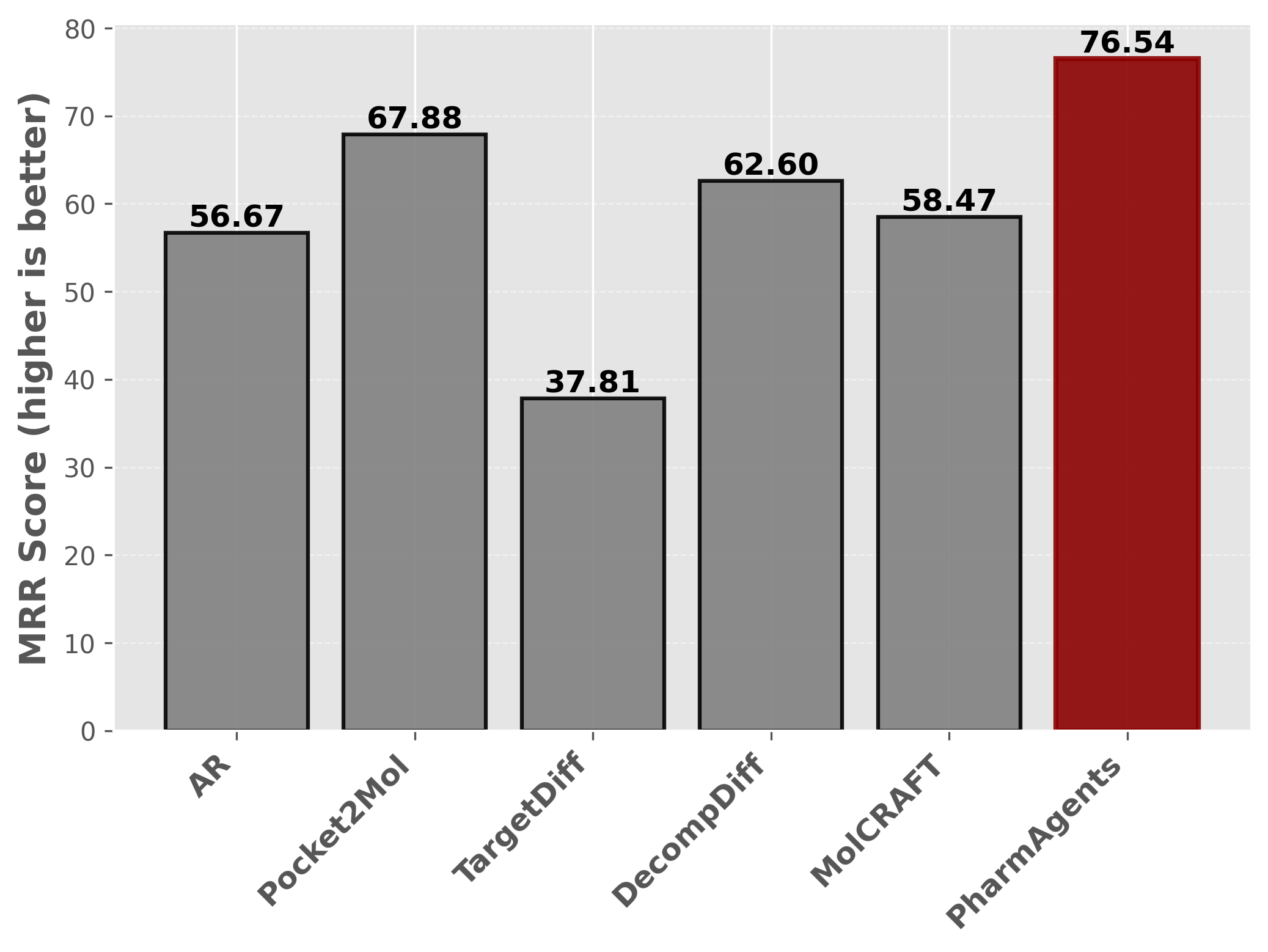}
        \caption{MRR Score (higher is better)}
    \end{subfigure}
    \hfill
    \begin{subfigure}[b]{0.3\textwidth}
        \centering
        \includegraphics[width=\textwidth]{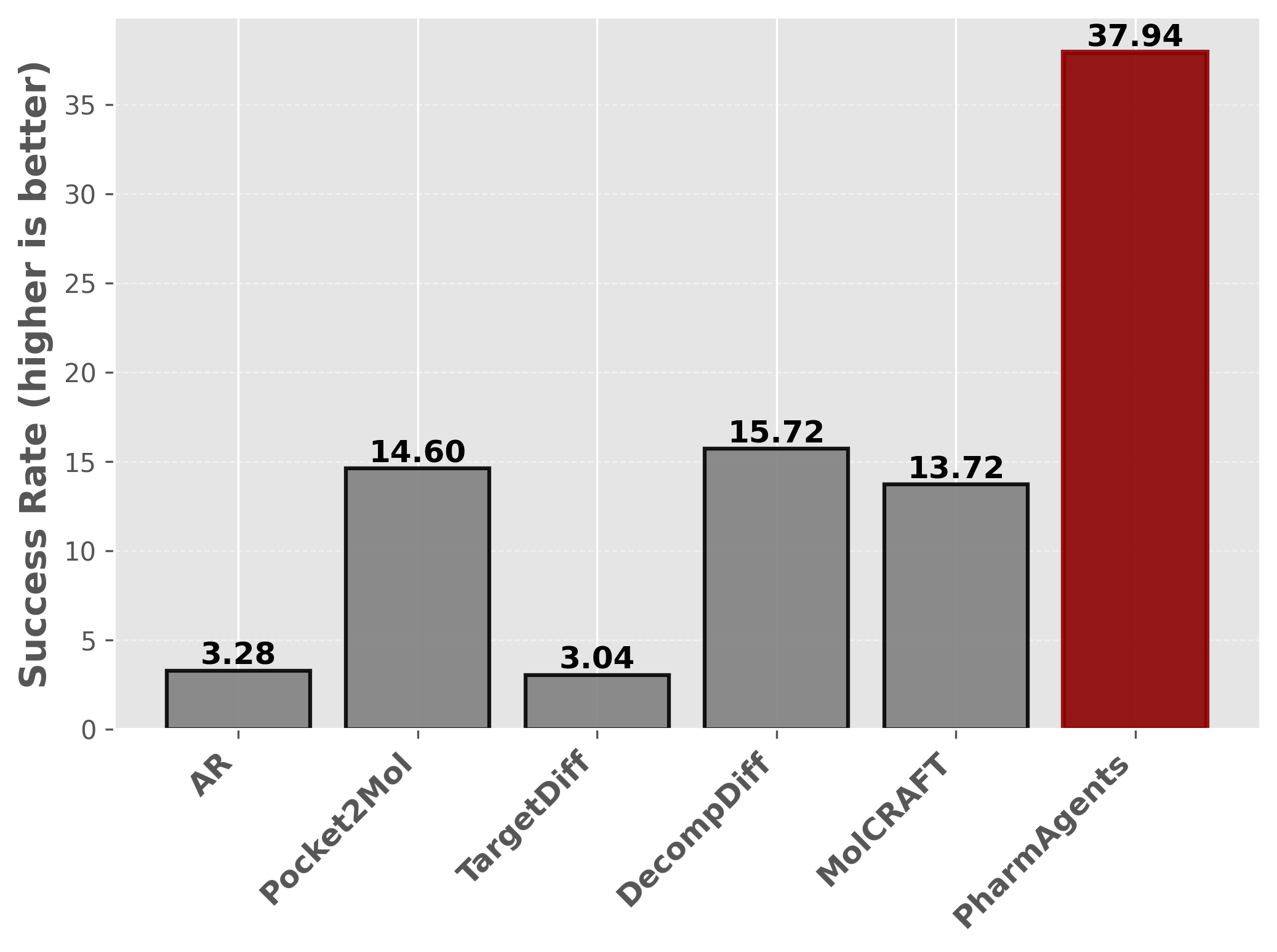}
        \caption{Success Rate (higher is better)}
    \end{subfigure}
    \hfill
    \begin{subfigure}[b]{0.3\textwidth}
        \centering
        \includegraphics[width=\textwidth]{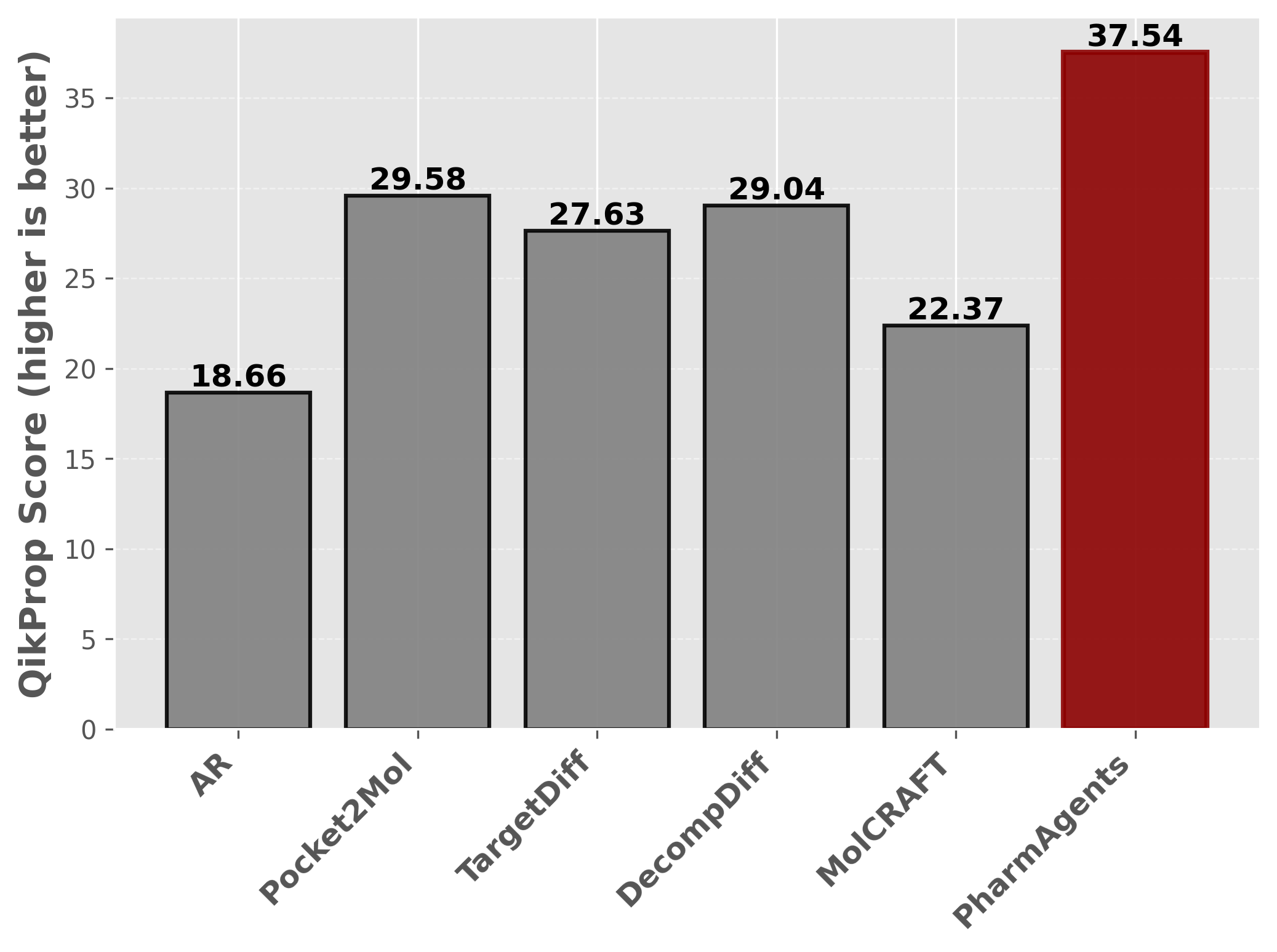}
        \caption{QikProp Score (higher is better)}
    \end{subfigure}

    \caption{Comparison of different methods across multiple metrics. PharmAgents is highlighted in \textcolor{red}{red}.}
    \label{fig:lead opt main results}
\end{figure}

To evaluate the overall generation capability of our LLM-powered Lead Identification and Optimization Module, we compared its performance against several state-of-the-art structure-based drug design models, including Pocket2Mol \cite{peng2022pocket2mol}, TargetDiff \cite{guan3d}, DecompDiff \cite{guan2023decompdiff}, and MolCraft \cite{qumolcraft}.

For assessment, we utilized commonly used metrics such as the Vina docking score to evaluate binding affinity and the SA score to measure synthesizability. Additionally, we incorporated drug-likeness metrics proposed by \cite{gao2025pushingboundariesstructurebaseddrug}, including the molecular reasonability ratio (MRR) and the QikProp properties pass ratio. Furthermore, we employed the success rate metric defined by \cite{gao2025pushingboundariesstructurebaseddrug}, which considers a molecule to be a viable lead only if it satisfies all property constraints. This comprehensive evaluation accounts not only for interaction strength (docking score) but also for synthesizability and molecular reasonability. All evaluations were conducted on the CROSSDOCKED dataset \cite{francoeur2020three}.

As shown in Figure \ref{fig:lead opt main results}, the Lead Identification/Optimization module in the PharmAgents system achieves better results compared to SOTA drug design models, especially for success rate, with a near 3X improvement.  

\begin{figure}[h]
\begin{center}
\centerline{\includegraphics[width=1\columnwidth]{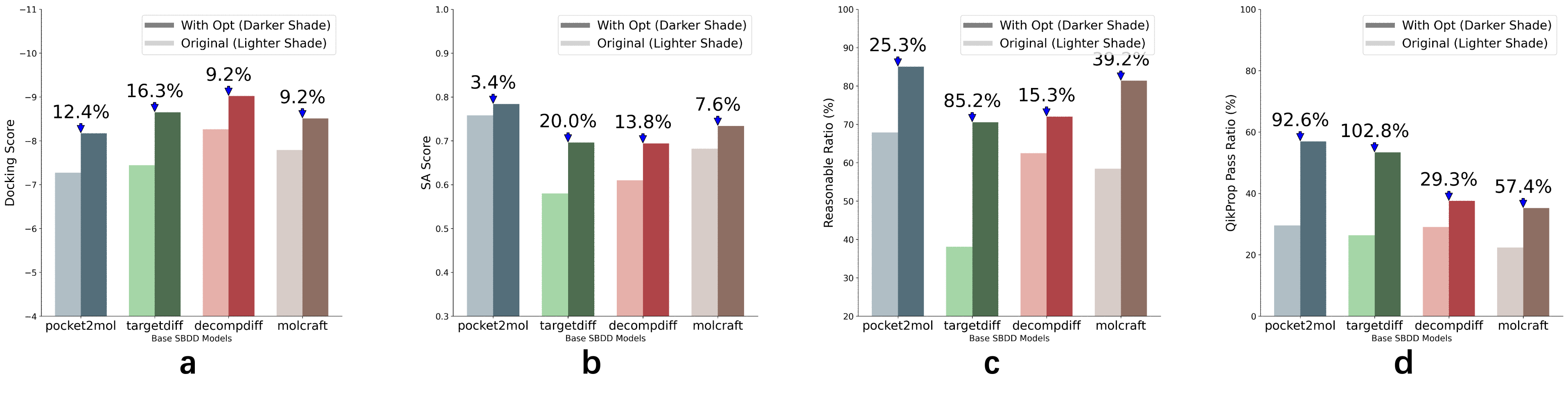}}
\caption{Metric-wise improvements over different SBDD baselines achieved by our module.}
\label{fig: lead_opt_results}
\end{center}
\end{figure}

As shown in Figure~\ref{fig: lead_opt_results}, our lead optimization module consistently improved performance across different metrics and molecule generation methods. Notably, we observed a maximum improvement of 16.3\% in docking scores, 85.2\% in the molecular reasonability ratio, 20\% in SA scores, and 102.8\% in the QikProp pass ratio. These results highlight the effectiveness of our lead optimization model in enhancing multiple molecular properties, including binding affinity, synthesizability, and drug-likeness.

\begin{figure}[h]
    \centering

    \begin{subfigure}[b]{0.245\textwidth}
        \centering
        \includegraphics[width=\textwidth]{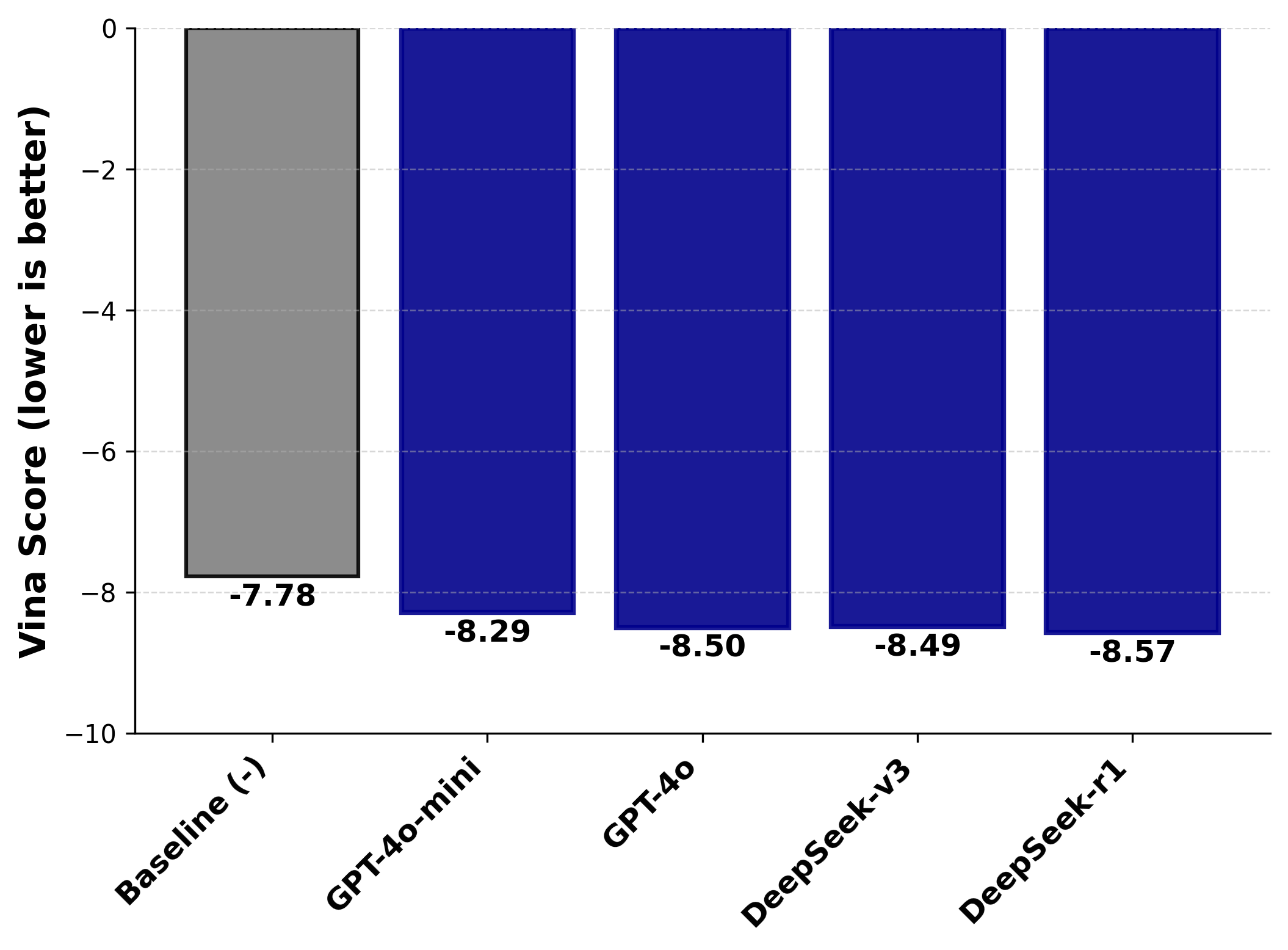}
        \caption{Vina Score}
    \end{subfigure}
    \hfill
    \begin{subfigure}[b]{0.245\textwidth}
        \centering
        \includegraphics[width=\textwidth]{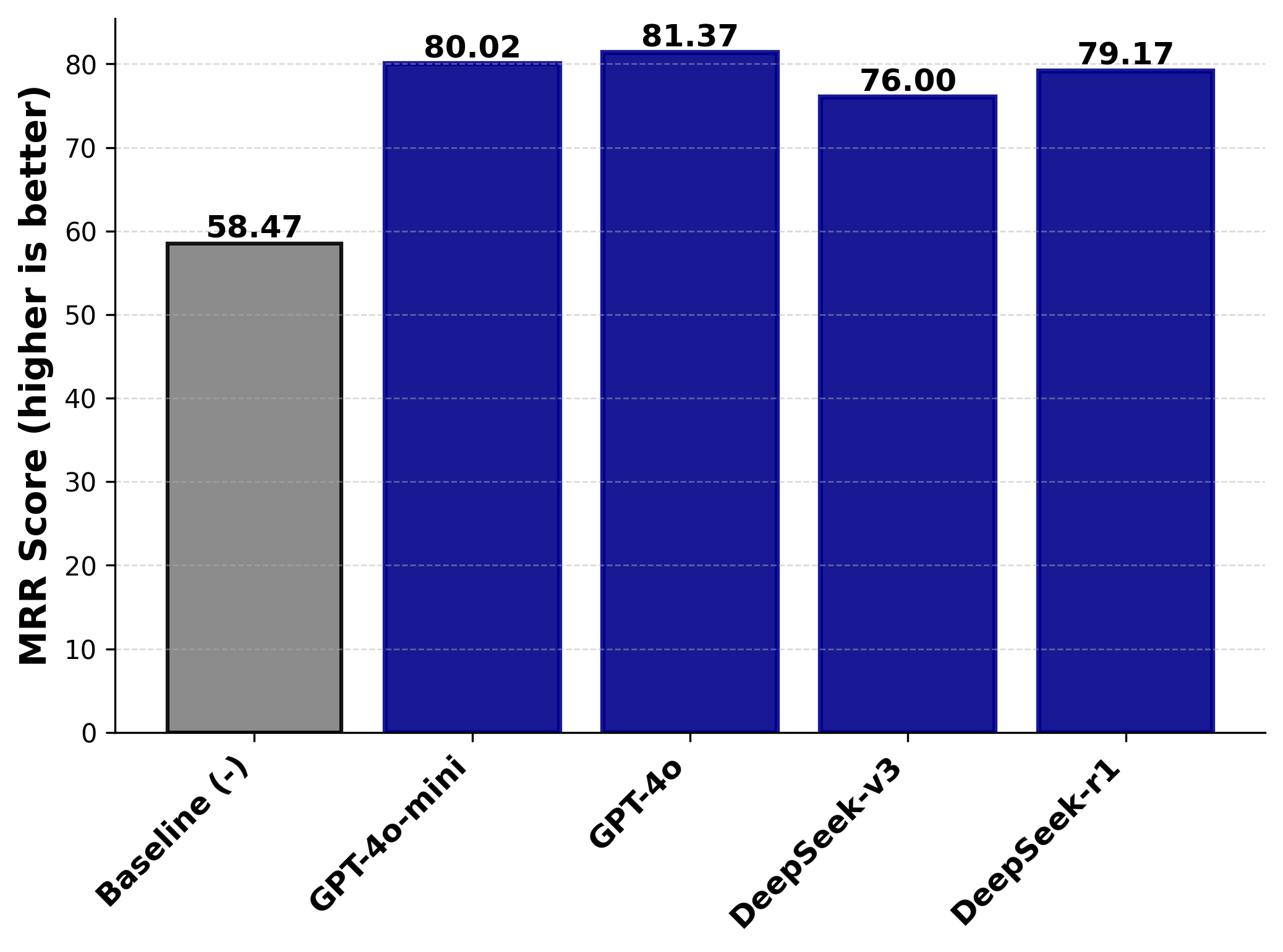}
        \caption{MRR Score}
    \end{subfigure}
    \hfill
    \begin{subfigure}[b]{0.245\textwidth}
        \centering
        \includegraphics[width=\textwidth]{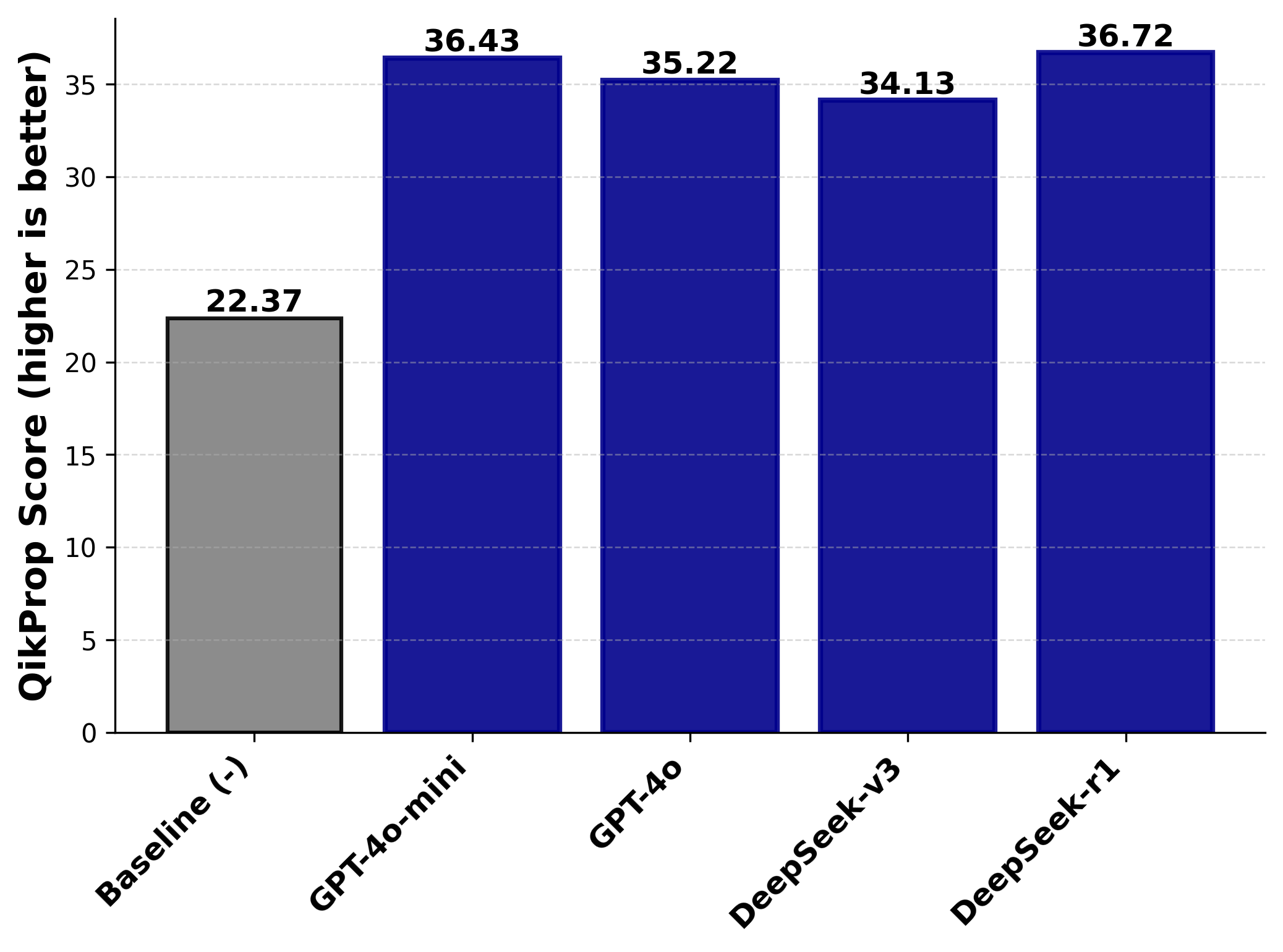}
        \caption{QikProp Score}
    \end{subfigure}
    \hfill
    \begin{subfigure}[b]{0.245\textwidth}
        \centering
        \includegraphics[width=\textwidth]{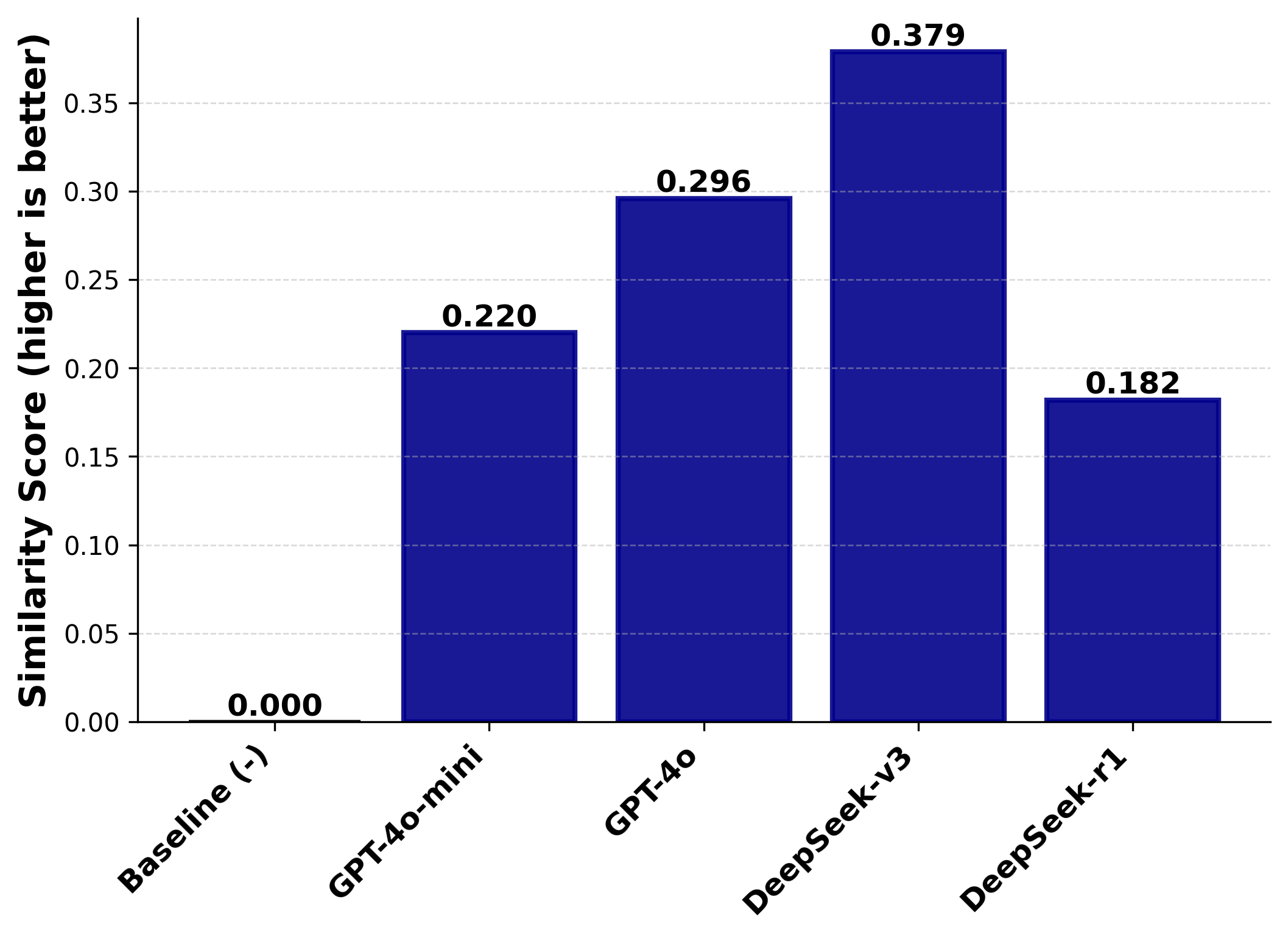}
        \caption{Similarity Score}
    \end{subfigure}

    \caption{Ablation study results across different LLMs. The baseline is in \textcolor{gray}{gray}, and LLM-based results are highlighted in \textcolor{blue}{blue}.}
    \label{fig:ablation_study}
\end{figure}

All results were obtained using GPT-4o \cite{achiam2023gpt} as the LLM. We also tested different LLMs, including GPT-4o-mini, GPT-4o, DeepSeek-V3 \cite{liu2024deepseek}, and DeepSeek-R1 \cite{guo2025deepseek}. The results are summarized in Figure~\ref{fig:ablation_study}. Molecules generated by GPT-4o-mini performed worse than other methods in terms of docking scores. Notably, DeepSeek-V3 required minimal modification to generate improved molecules, achieving the highest similarity to the initial molecules. In contrast, DeepSeek-R1 did not perform as well. This might be due to DeepSeek-R1 incorporating a more extensive reasoning process, which leads to a broader exploration of molecular modifications.

\subsubsection{PCC Evaluation}

\begin{figure}[h]
\begin{center}
\centerline{\includegraphics[width=1\columnwidth]{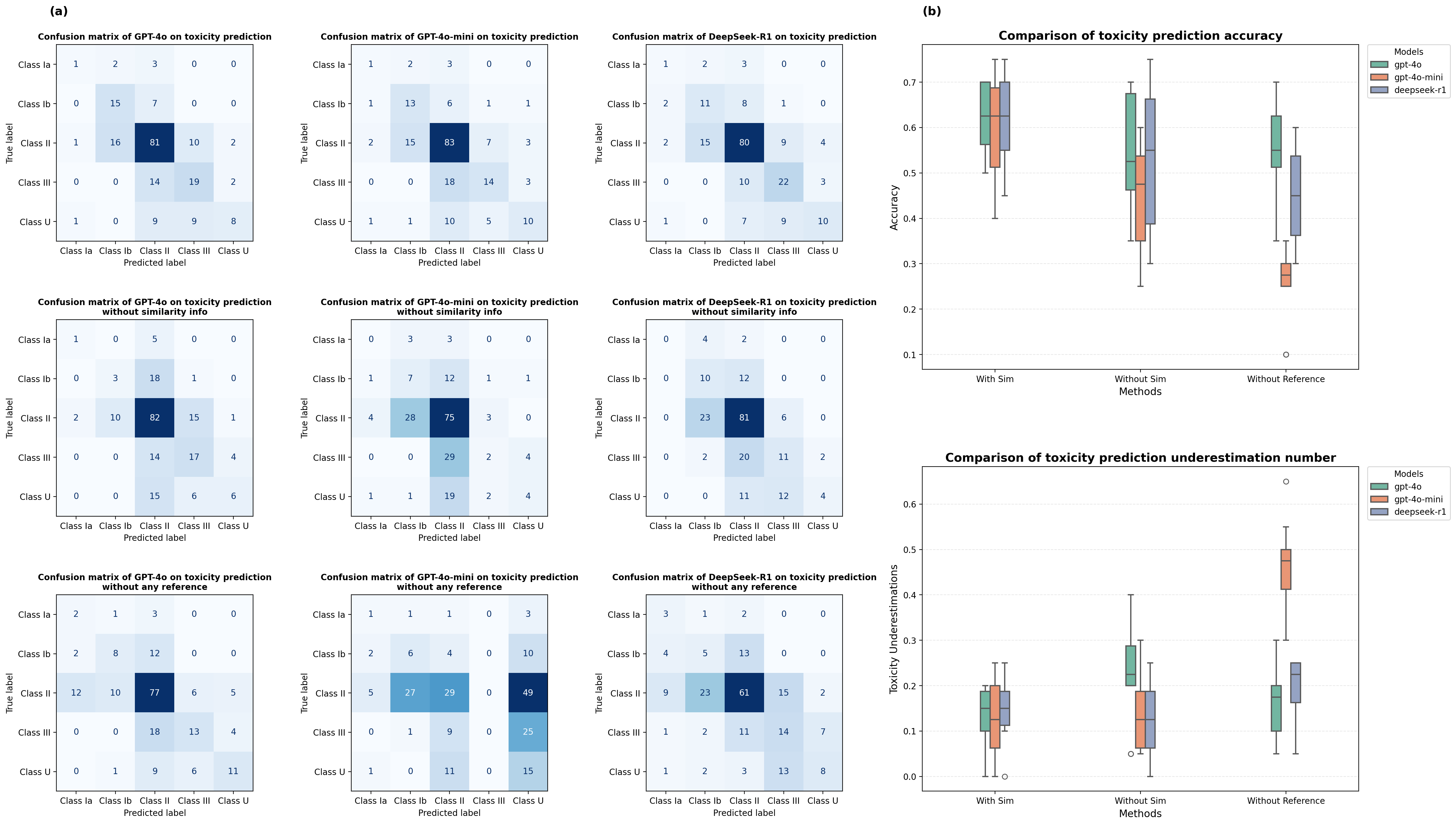}}
\caption{A comprehensive evaluation of toxicity prediction task on different models and methods}
\label{fig: tox comprehensive}
\end{center}
\end{figure}

\paragraph{Toxicity prediction}
For the Metabolism and Toxicity Assessment Agent, the main objective is to make sure the upstream molecules and their metabolites are less likely to be hazardous to human. Therefore, with the metabolism deduction covered by the leading deep learning model MetaTrans, the toxicity prediction task is landed on the LLM part of the agent, which relies highly on analogy, comparison and summarization.

To verify the ability of the agent on this task, we designed a five-category classification task which matched the WHO acute toxicity criteria. From the data aspect, we split the 9844-chemical toxicity dataset into a 9000-chemical reference dataset, and a 844-chemical test dataset, from which 200 chemicals are picked and sorted into 10 groups to test the agent's toxicity prediction ability. Method-wise, we developed two more different techniques of toxicity information transmission other than the aforementioned one which will be referred to as method \textit{With Sim} later in the analysis, to compare the influence on the prediction with different degree of information relevance. In the first method which we call \textit{Without Sim}, instead of finding the most similar molecules, we fill the context with as much chemical toxicity information as possible to let the LLM find its own answer among the vast dataset. And in the second method which we call \textit{Without Refer}, we leave the LLM with no reference from the dataset at all, relying only on the pre-defined toxicity knowledge and the inherent knowledge from the training corpus of the LLM. Finally, we have also chosen three prevalent models, GPT-4o, GPT-4o-mini and DeepSeek-R1, as the LLM part, representing normal LLM, distilled LLM and long-chain reasoning LLM respectively.

For the assessment metrics, we focused mainly on the prediction accuracy and the underestimations of toxicity that might result in adverse effects for human. 

As shown in Figure \ref{fig: tox comprehensive}-(a), all the combinations of different methods and models are tested and put into the confusion matrices, where the traces of the matrices indicate the correct prediction numbers, and the sums of the upper triangle areas (without the main diagonal) stand for the total cases of underestimation. Both metrics are gathered and rearranged into box-plots shown in Figure \ref{fig: tox comprehensive}-(b). It is clear from the test data that, firstly, the less information about the most similar molecules, the more likely the numbers in the confusion matrices are deviated from the diagonal, leading to lower accuracy and higher chance to underestimate the toxicity. Secondly, the results also demonstrate that, with larger model in the agent, the dependence on the relevant molecule information becomes lower, since larger models might carry more knowledge about chemicals and toxicity, which could be of great use in the prediction task. For smaller models like GPT-4o-mini, the deprivation of relevant knowledge can lead to problematic performances. Thirdly, from the comparison between GPT-4o and DeepSeek-R1, we can also tell that the long-chain reasoning might slightly undermine the model's accuracy as well, and we suspect it is due to the overthinking issue from the long-chain reasoning.

In conclusion, GPT-4o is deemed to be the best model for this task, as it shows high accuracy for toxicity prediction, and relatively low underestimation likelihood. Moreover, it can also maintain high robustness in different levels of information adequacy.

\begin{figure}[!htbp]
\begin{center}
\centerline{\includegraphics[width=1.0\columnwidth]{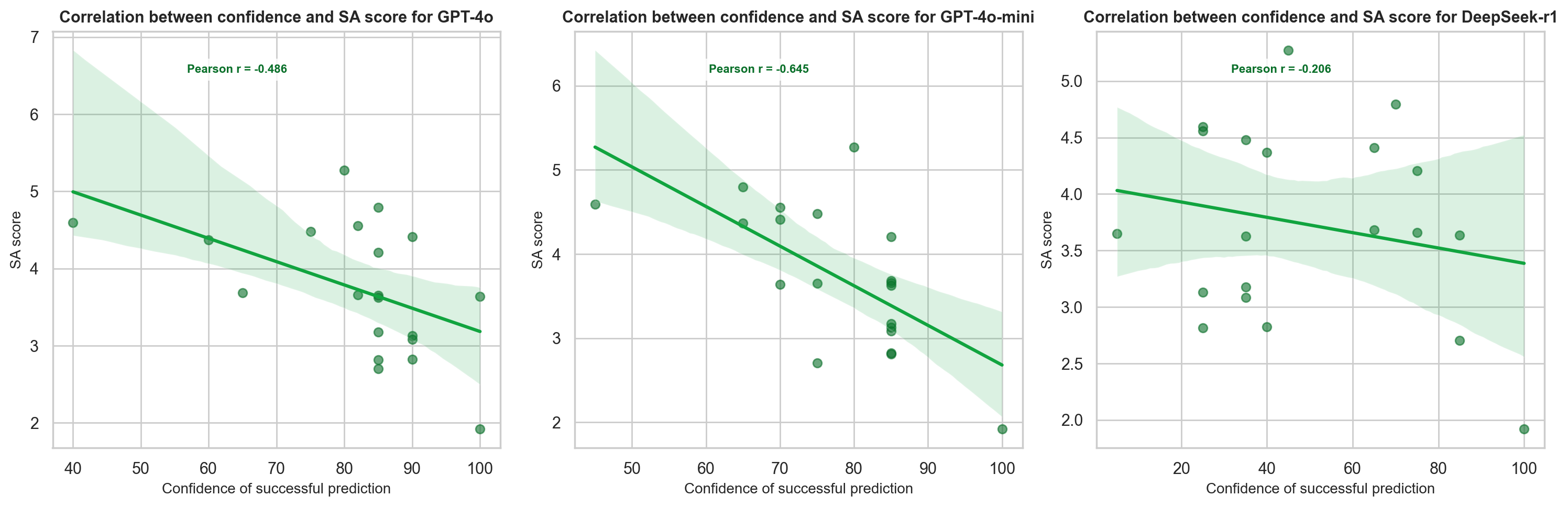}}
\caption{Correlation of confidence of successful synthesis and SA score on different models}
\label{fig: SA correlation}
\end{center}
\end{figure}

\paragraph{Synthesizability analysis}
To test the Synthesis Assessment Agent, we mainly researched on the correlation between the SA score of chemicals and the synthesis success confidence given by the LLM.

We chose 20 molecules which were generated from the SBDD models and then modified by the Lead Optimization module as the test set, and let them undergo the process in the Synthesis Assessment Agent to yield a confidence rating for each of them, to finally compare the ratings with their SA scores using the Pearson correlation coefficient. In this test, the same three models as above were used.

From the linear fitting graphs shown in Figure \ref{fig: SA correlation}, we can see that the results of GPT-4o-mini present the most obvious negative correlation with the Pearson coefficient at -0.645. Meanwhile, the results from GPT-4o and DeepSeek-R1 are relatively lower, with the coefficient value of -0.486 and -0.206 accordingly. To further analyze the reasons behind the ratings, we also did a human evaluation on the molecules and the deduction process of the LLMs, in which we found that sometimes larger models, for instance, DeepSeek-R1, tend to give low score for molecules that should have been easier to synthesize. From the narrations of the LLMs, we found that the low scores were caused by the issue that the LLMs had been too much aware of the missing moieties of the molecules which hadn't been included by the starting molecules, therefore regarding the whole synthetic process as incomplete. Yet for the smaller GPT-4o-mini, it stuck with the rules in the retrosynthesis knowledge in a better fashion and avoided much unnecessary overthinking, although sometimes giving repetitive answers for product molecules with various SA scores, its answers all made some sense combined with the synthesis path generated by UAlign. 

Due to the fact that sometimes SA score itself might not be enough to judge the synthetic availability of a molecule, we deem the analysis from GPT-4o-mini a convincing interpretation and addition to the SA score, since it is not 100\% related to SA score, yet still retaining high correlation and explainability. And for models similar to DeepSeek-R1, they are, to some extent, less compatible for this task, but they have still shown reasonable and relatively accurate understandings in chemistry and synthetic domains nonetheless.

\section{Beneficial and Future Potential of Virtual Pharma}

\begin{figure}[!htbp]
\begin{center}
\centerline{\includegraphics[width=1.0\columnwidth]{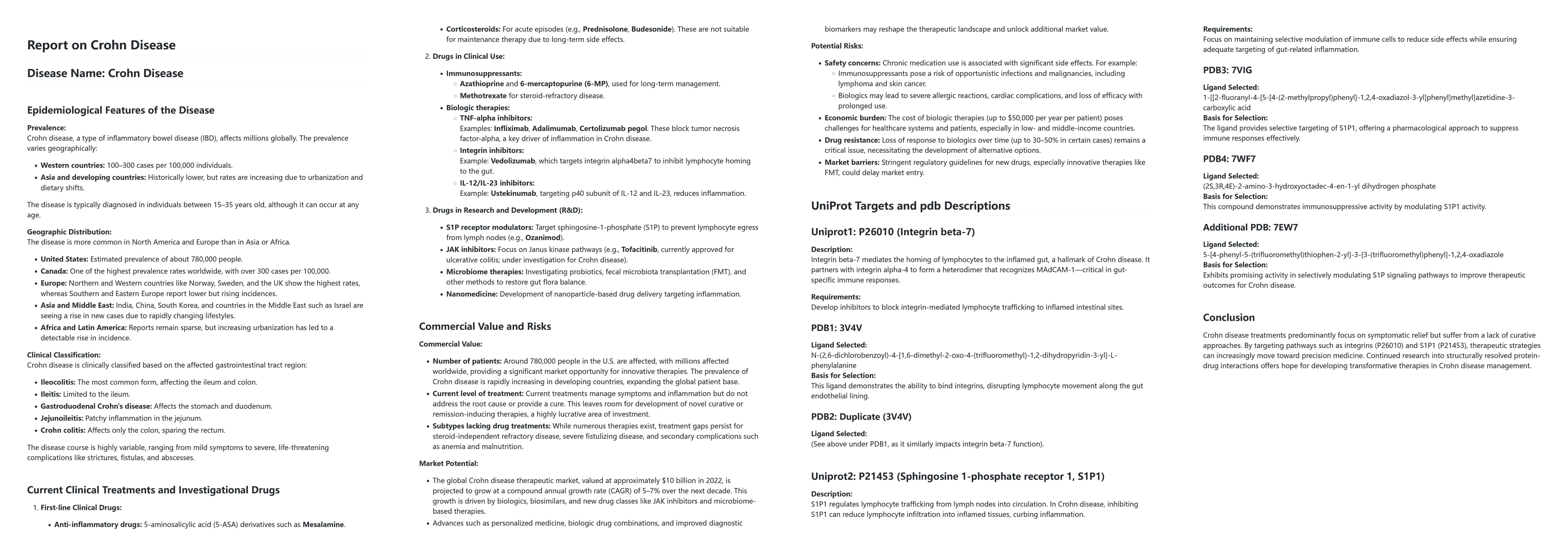}}
\caption{An example of report produced by \textbf{Research Expert Agent}}
\label{fig: report_draft}
\end{center}
\end{figure}

\begin{figure}[!htbp]
\begin{center}
\centerline{\includegraphics[width=1.0\columnwidth]{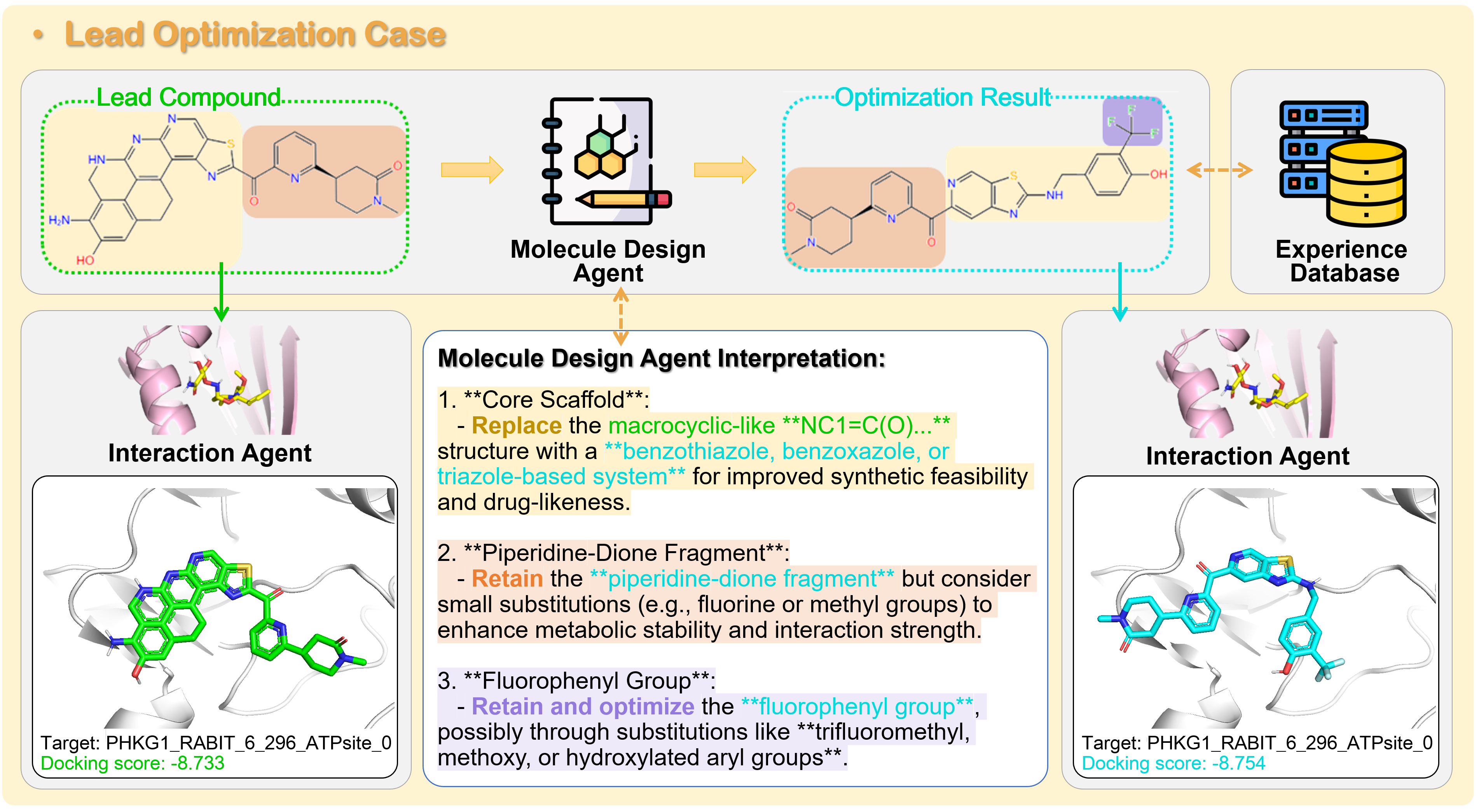}}
\caption{Case study showcasing the Lead Optimization module’s molecular design process and the interpretation of designed structures in the generated molecules.}
\label{fig: cidd case}
\end{center}
\end{figure}

\subsection{Interpretability}

The key advantage of PharmAgents over other opaque systems that rely solely on machine-learning models or computational tools is its interpretability. Every decision and output from each module is accompanied by reasoning generated by an LLM. As demonstrated in the previous section, these explanations can be directly linked to module outputs, ensuring transparency throughout the process. 

Throughout the target discovery process, a research expert agent records each step and produces a final report. One of the reports is shown in Figure \ref{fig: report_draft}. The report not only includes existing information of the disease and its market value, but also provides an analysis of the selected targets and the reasons for the selection.

As shown in Figure \ref{fig: cidd case}, the designs generated by the design agent can be traced back to specific molecular fragments, with the reasoning behind each modification clearly reflected in the altered structure. This ensures that the entire design process is fully explainable, allowing users to not only assess the reliability of the results but also evaluate whether the designed molecule effectively realizes its intended purpose. This interpretability is particularly valuable in the drug discovery pipeline, as it enhances trust in the system and supports more informed decision-making.

\subsection{Evolvement Potential}
\label{sec: evolvement}

\begin{figure*}[!htbp]
    \centering
    \begin{subfigure}[b]{0.42\textwidth}
        \centering
        \includegraphics[width=\textwidth]{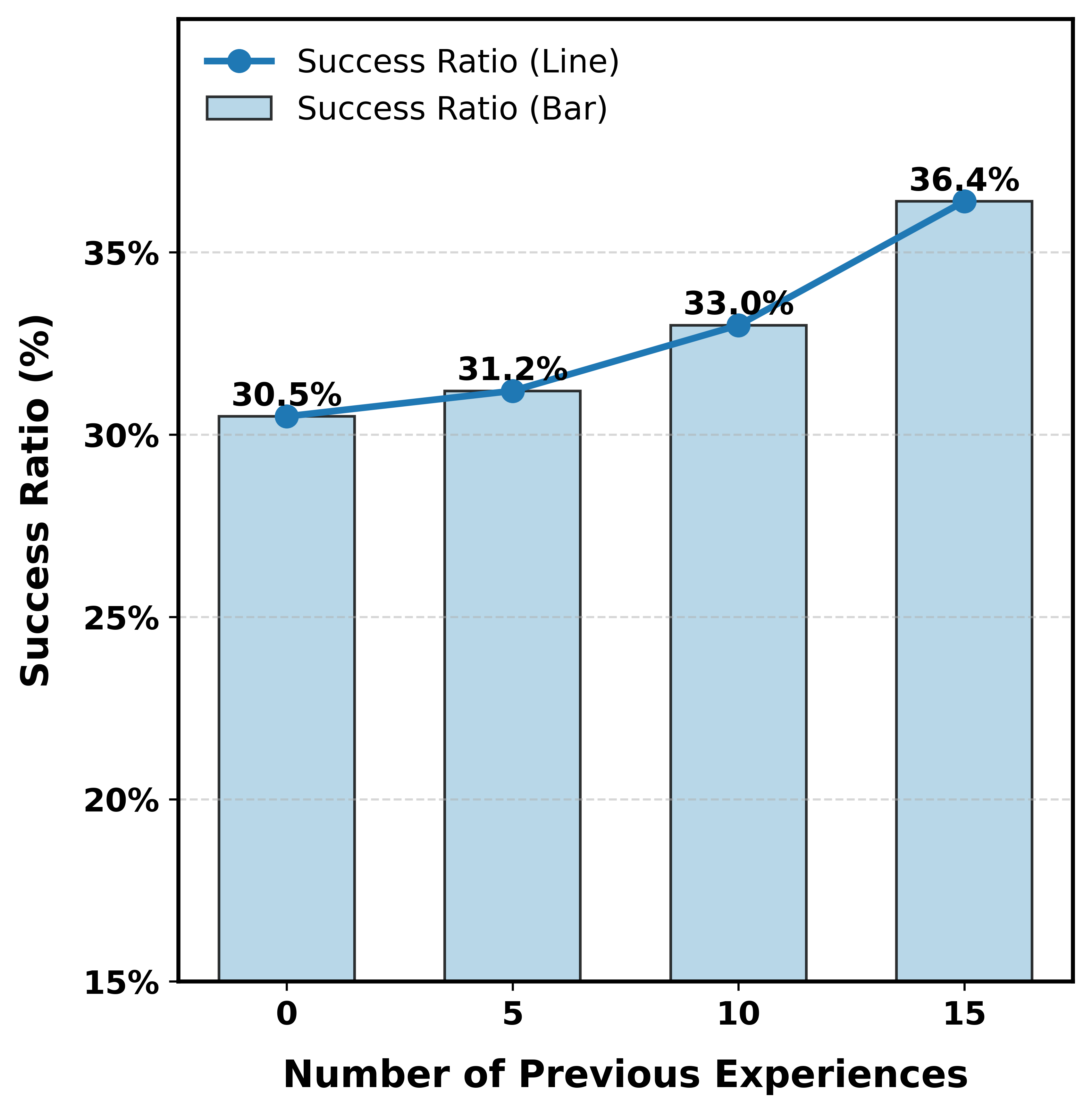} 
        \caption{}
        \label{fig: icl}
    \end{subfigure}
    \hfill
    \begin{subfigure}[b]{0.42\textwidth}
        \centering
        \includegraphics[width=\textwidth]{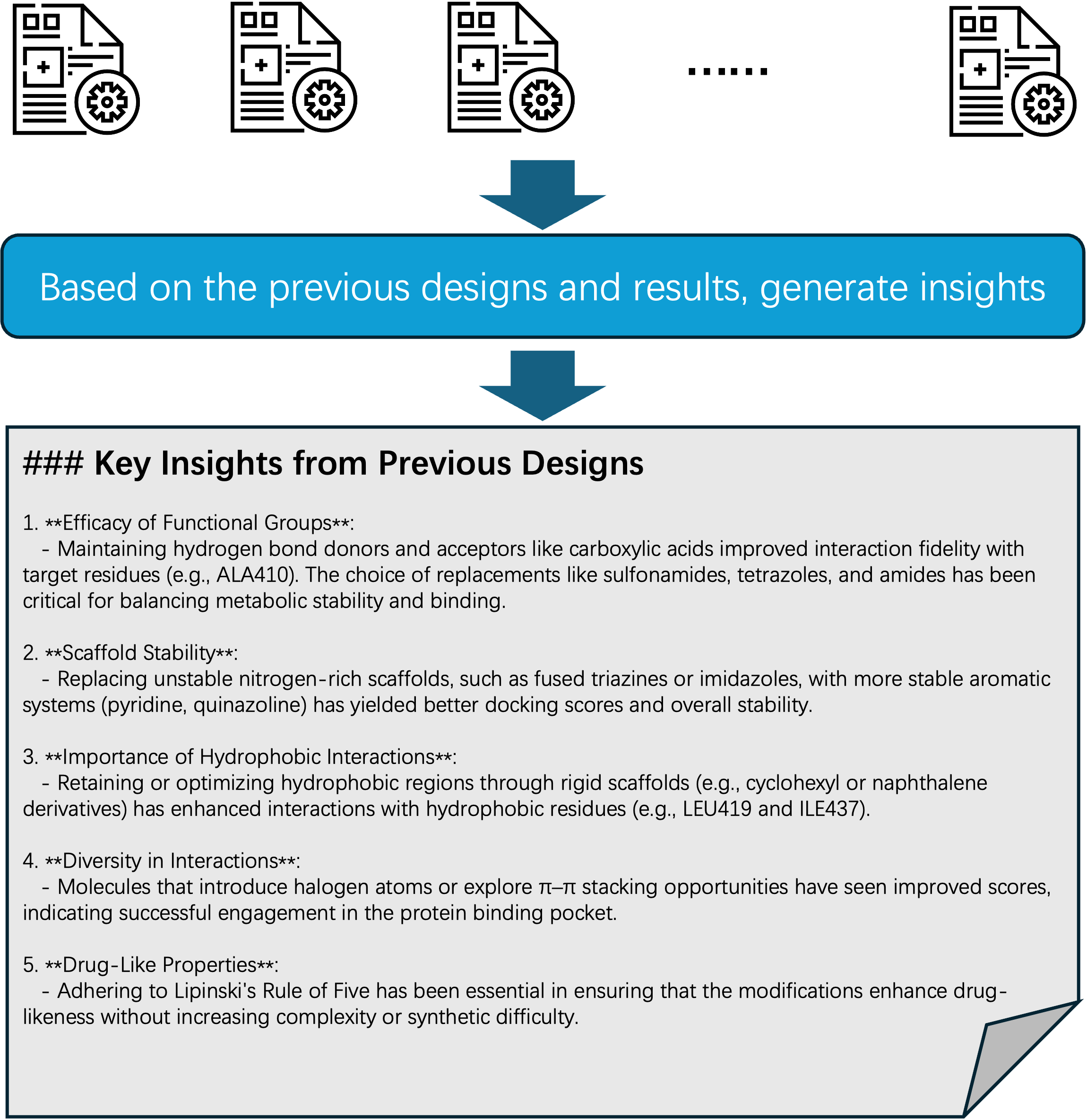} 
        \caption{}
        \label{fig: exp}
    \end{subfigure}
    \caption{PharmAgents demonstrates self-evolution through accumulated experience. \textbf{(a)} The success rate of generated molecules improves as the system gains more prior experience. \textbf{(b)} The design agent leverages past experience to generate valuable insights for molecular design.}
    \label{fig:twofigs}
\end{figure*}

Another advantage of an LLM-powered system is its ability to interact with users in real-time. Unlike traditional machine learning models, which require retraining to update weights based on feedback, LLMs can immediately incorporate feedbacks. Moreover, these interactions serve as experiential learning, enabling the system to improve through in-context learning for future drug discovery attempts. This allows for rapid error correction with expert guidance, fostering seamless collaboration between Virtual Pharma and human specialists. As a result, the drug discovery pipeline becomes more efficient, leveraging both speed and accuracy to accelerate breakthroughs. In the future, we will continue to build the interaction ability of the PharmAgents system, to make it a fully interactive server that can be used by Pharmaceutical Scientists to accelerate the drug discovery procedure.

The use of LLM-based agents is particularly beneficial due to their ability to \textbf{self-evolve}. In a real-world virtual pharma environment, vast amounts of experimental data are accumulated for each target, guiding future drug designs. Similarly, in our Virtual Pharma, past molecular designs and corresponding reports serve as a foundation for the continuous improvement of design agents, enabling them to generate increasingly optimized molecules.

To validate this concept, we conducted an experiment where previous design experiences were integrated into the design agent to derive insights, which were then leveraged for subsequent designs. As illustrated in Figure \ref{fig: icl}, an increase in the number of prior experiences correlates with a higher overall success rate. Additionally, Figure \ref{fig: exp} demonstrates that the insights generated by the design agent are logical and meaningful. Those insights can also be useful assets for human experts to use and help with their work. These results indicate that our system has the potential for continuous self-improvement, evolving as more designs are generated and more experience is accumulated.

\subsection{Drug Lifecycle Management Potential}

In this study, we have developed a fully automated framework capable of generating preclinical candidates from disease name inputs, marking a significant advancement in the early stages of drug discovery. Looking forward, we aim to extend the capabilities of our PharmAgent framework beyond the identification of novel chemical entities by exploring the integration of multi-agent systems into the broader drug development pipeline.

While early-stage drug discovery is critical, subsequent processes—including clinical trials, regulatory approval, and post-marketing surveillance—are essential to ensuring both therapeutic efficacy and patient safety. These stages, however, are often time-consuming, involving extensive documentation and the analysis of complex clinical response data. Given recent advancements in LLMs and AI-driven automation, there is substantial potential to streamline these tasks while maintaining regulatory rigor. As illustrated in Figure~\ref{fig: virtual}, the time required for these later-stage processes frequently exceeds that of the initial discovery phase. Reducing these timelines with LLMs or agents could accelerate patient access to life-saving treatments and enhance the economic viability of drug development. Future work will explore the application of LLM-powered agents in automating administrative and analytical aspects of the drug lifecycle.

\section{Conclusion}

In this paper, we introduce PharmAgents, a novel system that simulates the entire drug discovery pipeline using a multi-agent framework powered by LLMs. Each agent performs a specific task in the drug development process, collaborating seamlessly from target discovery to lead identification, optimization, and preclinical evaluation. Our results demonstrate that PharmAgents effectively executes each stage with interpretable reasoning, enabling not only efficient but also explainable and transparent automated drug discovery. In the lead generation and optimization phase, PharmAgents outperforms state-of-the-art molecular generation models by improving key design metrics and tripling the success rate. Furthermore, it enhances interpretability by providing rational justifications for each molecular modification. In our experiments, it has been proved that LLMs can uncover implicit knowledge associations from vast amounts of pretraining data. The system also exhibits self-evolution, continuously learning from past design experiences to refine future drug candidates. This work establishes a new paradigm in AI-driven drug discovery, leveraging LLM-powered multi-agent systems to create a virtual pharmaceutical ecosystem. With ongoing advancements in interpretability, interaction capabilities, self-improvement, and Drug Lifecycle Management, PharmAgents has the potential to revolutionize the drug discovery process, making it more efficient, transparent, and intelligent.


\clearpage  
\newpage

\bibliography{sample}

\end{document}